\documentclass[12pt,a4paper]{article}
\usepackage{a4wide}
\usepackage{epsfig}
\usepackage{cite}
\usepackage{amsmath}
\usepackage{graphicx}
\usepackage{latexsym}
\usepackage{array}

\newlength{\absize}
\newcommand{\dd}{\mbox{{\rm d}}}
\newcommand{\half}{{\textstyle\frac{1}{2}}}

\def\lsim{\mathrel{\rlap{\raise 2.5pt \hbox{$<$}}\lower 2.5pt
\hbox{$\sim$}}}

\renewcommand{\Re}{\mbox{Re}}

\newcommand{\Lumint}{{\cal L}_{\rm int}}
\newcommand{\rpv}{{\not\!\! R_p}}

\begin{document}

\thispagestyle{empty}
\renewcommand{\thefootnote}{\fnsymbol{footnote}}
\newpage\normalsize
\pagestyle{plain}
\setlength{\baselineskip}{4ex}\par
\setcounter{footnote}{0}
\renewcommand{\thefootnote}{\arabic{footnote}}
\newcommand{\preprint}[1]{%
\begin{flushright}
\setlength{\baselineskip}{3ex} #1
\end{flushright}}
\renewcommand{\title}[1]{%
\begin{center}
\LARGE #1
\end{center}\par}
\renewcommand{\author}[1]{%
\vspace{2ex}
{\Large
\begin{center}
 \setlength{\baselineskip}{3ex} #1 \par
\end{center}}}
\renewcommand{\thanks}[1]{\footnote{#1}}
\renewcommand{\abstract}[1]{%
\vspace{2ex}
\normalsize
\begin{center}
\centerline{\bf Abstract}\par
\vspace{2ex}
\parbox{\absize}{#1\setlength{\baselineskip}{2.5ex}\par}
\end{center}}

\vspace*{4mm} 

\title{Spin-identification of the Randall-Sundrum resonance in
lepton-pair production at the LHC } \vfill

\author{P. Osland,$^{a,}$\footnote{E-mail: per.osland@ift.uib.no}
A. A. Pankov,$^{b,}$\footnote{E-mail: pankov@ictp.it} 
N. Paver$^{c,}$\footnote{E-mail: nello.paver@ts.infn.it} and 
A. V. Tsytrinov$^{b,}$\footnote{E-mail: tsytrin@rambler.ru}}

\begin{center}
$^{a}$Department of Physics and Technology, University of Bergen,
Postboks 7803, N-5020  Bergen, Norway\\
$^{b}$The Abdus Salam ICTP Affiliated Centre, Technical
University of Gomel, 246746 Gomel, Belarus\\
$^{c}$University of Trieste and INFN-Trieste Section, 34100
Trieste, Italy
\end{center}
%
%
%
\vfill

\abstract{ The determination of the spin of the quantum states exchanged in
the various non-standard interactions is a relevant aspect in the
identification of the corresponding scenarios.  We discuss the {\it
identification} reach at LHC on the spin-2 of the lowest-lying Randall-Sundrum
resonance, predicted by gravity with one warped extra dimension, against
spin-1 and spin-0 non-standard exchanges with the same mass and producing the
same number of events in the cross section. We focus on the angular
distributions of leptons produced in the Drell-Yan process at the LHC, in
particular we use as basic observable a ``normalized'' integrated angular
asymmetry $A_{\rm CE}$. Our finding is that the 95\% C.L.\ {\it identification}
reach on the spin-2 of the RS resonance (equivalently, the exclusion reach on
both the spin-1 and spin-0 hypotheses for the peak) is up to a resonance mass
scale of the order of 1.0 or 1.6~TeV in the case of weak coupling between
graviton excitations and SM particles ($k/{\bar M}_{Pl}=0.01$) and 2.4 or
3.2~TeV for larger coupling constant ($k/{\bar M}_{Pl}=0.1$) for a
time-integrated LHC luminosity of 10 or $100\, {\rm fb}^{-1}$, respectively.
Also, some comments are given on the complementary r\^oles of the angular
analysis and the eventual discovery of the predicted second graviton
excitation in the identification of the RS scenario.}

\vspace*{20mm} \setcounter{footnote}{0} \vfill

\newpage
\setcounter{footnote}{0}
\renewcommand{\thefootnote}{\arabic{footnote}}

\section{Introduction} \label{sect:introduction}
A common feature of the different New Physics (NP) scenarios that go beyond
the Standard Model (SM) is the predicted existence of heavy new particles or
``resonances'', that can be either produced or exchanged in reactions studied
at high energy colliders. Such non-standard objects are expected to be in the
TeV mass range, and could be revealed directly as peaks in the energy
dependence of the measured cross sections.

\par For any model, given the expected statistics and experimental
uncertainties, one can assess the corresponding {\it discovery} reach by
determining the upper limit of the mass range where the resonance signal can
be detected above the SM cross section to a given confidence level.

\par On the other hand, once a peak in the cross section is observed, further
analysis is needed to distinguish the underlying non-standard dynamics against
the other scenarios that potentially may cause a similar effect. In this
regard, the expected {\it identification} reach is defined as the upper limit
of the mass range where the model could be identified as the source of the
peak or, equivalently, the other competitor models can be excluded for all
values of their parameters. The determination of the spin of the resonance
represents therefore an important selection among different classes of
non-standard interactions.

\par Here, we consider the discrimination reach on the lowest-lying spin-2
Randall--Sundrum (RS) graviton resonance \cite{Randall:1999ee}, that could be
obtained from measurements of the Drell--Yan (DY) lepton-pair production
processes ($l=e,\mu$) at the LHC:
\begin{equation}
p+p\to l^+l^-+X. \label{proc}
\end{equation}

\par The determination of the spin-2 of the lowest-lying RS graviton resonance
exchange against the spin-1 hypothesis in the context of experiments at LHC
has recently been discussed, e.g., in
Refs.~\cite{Allanach:2000nr,Allanach:2002gn,Cousins:2005pq,Belotelov:2006bh},
and some attention to the case of the spin-0 hypothesis has been given in
Ref.~\cite{Cousins:2005pq}. 
Also, the exclusion of the spin-2 hypothesis against the spin-1
Stueckelberg $Z^\prime$ was discussed in Ref.~\cite{Feldman:2006wb}.
An experimental search for spin-2, spin-1 and
spin-0 new particles decaying to DY dilepton pairs has recently been performed
at the Fermilab Tevatron $p\bar p$ collider~\cite{Abulencia:2005nf}.

\par We would like to complement those analyses and assess the extent to which
the domain in the RS parameters allowed by the discovery reach on the
resonance is reduced by the request of simultaneous exclusion of {\it both}
the hypotheses of spin-1 and spin-0 exchanges with the same mass, and
mimicking the same peak in the dilepton invariant mass distribution (same
number of events).

\par As is well known, the main tool to differentiate among the spin exchanges
in the process (\ref{proc}) uses the different, and characteristic,
dependencies on the angle $\theta_\text{cm}$ between the incident quark or
gluon and the final lepton in the dilepton center-of-mass frame. We shall base
our discussion on the integrated center-edge asymmetry $A_{\rm CE}$, that has
the property of directly disentangling the spin-2 from vector interactions as
illustrated in Refs.~\cite{Dvergsnes:2004tw,Osland:2003fn}. This method
represents an alternative to the use of the differential distributions ${\rm
d}N/{\rm d}\cos\theta_\text{cm}$, where $N$ represents the number of events.

\par We believe our analysis has sufficiently general features to be
applicable also to the identification of other spin-2 exchange interactions,
besides the RS model. We nevertheless prefer in the sequel to refer and expose
in detail the procedure in the case of that mentioned scenario. Moreover,
although strictly not necessary, we shall make comparisons with specific,
``physically motivated'' representative spin-0 and spin-1 models.

\par As an example of spin-0 contribution to the process (\ref{proc}), we can
consider the sneutrino exchange envisaged by supersymmetric theories with
$R$-parity breaking ($\rpv$) \cite{Kalinowski:1997bc,Allanach:1999bf}. In
$\rpv$ it is possible that some sparticles can be produced as $s$-channel
resonances, thus appearing as peaks in the dilepton invariant mass
distribution, if kinematically allowed.

\par Examples of competitor spin-1 mediated interactions, that can contribute
to process (\ref{proc}) and show up as peaks in the cross section are, besides
the SM $\gamma$ and $Z$, the heavy $Z^\prime$ exchanges \cite{Hewett:1988xc},
and we will refer to those models for the comparison with the RS resonance
exchange.

\par In Sec.~\ref{sect:crossstatistics} we discuss the LHC cross section and
statistics for the production of a Randall--Sundrum heavy graviton. In
Sec.~\ref{sect:sneutrino} we identify the ranges in the number of events and
mass that can originate from either the RS graviton or a spin-0 object, such
as a sneutrino. This common range is referred to as the ``signature space'' of
these models. A corresponding discussion is presented in
Sec.~\ref{sect:zprime} for the RS graviton and a spin-1 object, such as a
$Z^\prime$. In Sec.~\ref{sect:angdist} we review the relevant angular
distributions, and in Sec.~\ref{sect:gravitonidentification} we show how these
can be used to identify the RS graviton by means of the asymmetry $A_{\rm CE}$.
The results are collected in Sec.~\ref{sect:results},
where regions are identified in the plane spanned by the coupling strength
$c=k/{\bar M}_{Pl}$ and the mass $M_R$, where spin identification is
possible. Finally, Sec.~\ref{sect:conclusions} presents some concluding
remarks.

\section{LHC cross sections and statistics for RS} 
\label{sect:crossstatistics}
\par Considering its great popularity in the context of models solving the
gauge hierarchy problem, we here just recall that the simplest RS scenario is
based on one compactified warped extra spatial dimension and two branes, such
that the SM particles are confined to the so-called TeV brane while gravity
can propagate in the whole 5-dimensional space. In this scenario TeV-scale,
spin-2, narrow graviton resonances are predicted. The model depends on two
independent parameters, that can be chosen as the dimensionless ratio
$c=k/{\bar M}_{Pl}$, with $k$ the 5-dimensional scalar curvature and ${\bar
M}_{Pl}$ the reduced 4-dimensional Planck scale ($\bar{M}_{Pl}=M_{Pl}/\sqrt{8
\pi}$), and $m_1$, the mass of the lowest-lying graviton resonance. The masses
of the higher excitations $G^{(n)}$ are given by $m_n=m_1\, x_n/x_1$, where
$x_n$ are roots of the Bessel function $J_1(x_n)=0$ ($x_1=3.8317$,
$x_2=7.0156$, $x_3=10.1735$,..), and are therefore unevenly spaced. The mass
pattern may therefore be distinctive of the model, if higher excitations in
addition to the ground state would be discovered. A correlated parameter is
represented by the physical scale on the TeV brane $\Lambda_\pi=m_1/(c\,x_1)$,
whose inverse controls the strength of the graviton resonance coupling to
standard matter. The (theoretically) natural ranges for these parameters are
$0.01\leq c\leq 0.1$ and $\Lambda_\pi<10\, {\rm TeV}$
\cite{Davoudiasl:2000jd}.  Current discovery limits at 95\% C.L. from the
Fermilab Tevatron collider are for the first graviton mass: 300 GeV for
$c=0.01$ and 900 GeV for $c=0.1$ \cite{Abazov:2007ra}.
\subsection{Cross sections} 
\label{sect:crosssections}
In the SM, lepton pairs at hadron colliders can be produced at
tree level via the following parton-level processes:
\begin{equation}\label{Eq:qqbar-SM}
q\bar q \to \gamma,Z \to l^+ l^-.
\end{equation}
The first massive graviton mode $G^{(1)}$ of the RS model, in the sequel
denoted simply as $G$ (and the mass $m_1\equiv M_G$), can be produced via
quark--antiquark annihilation as well as gluon--gluon fusion,
\begin{equation} \label{Eq:qqbar-gg}
q\bar q \to G \to l^+ l^- \qquad \text{and} \qquad gg \to G \to
l^+ l^-,
\end{equation}
and can be observed as a peak in the dilepton invariant mass distribution. The
inclusive differential cross section for $G$ production and subsequent decay
into lepton pairs at the LHC can be expressed as the sum:
\begin{equation}
\frac{{\rm d}\sigma}{{\rm d}M\,{\rm d}y\,{\rm d}z}
=\frac{{\rm d}\sigma_{q \bar q}}{{\rm d}M\,{\rm d}y\,{\rm d}z} 
+\frac{{\rm d}\sigma_{gg}}{{\rm d}M\,{\rm d}y\,{\rm d}z},
\label{sigminclusive}
\end{equation}
where $M$ and $y$ are invariant mass and rapidity of the lepton pairs,
respectively, and $z=\cos\theta_\text{cm}$ with $\theta_\text{cm}$ the
lepton-proton angle in the dilepton center-of-mass frame.
Explicitly:
\begin{align}
\label{Eq:dsigma-dMdydz} &\begin{aligned} \frac{{\rm d}\sigma_{q \bar
q}}{{\rm d}M\,{\rm d}y\,{\rm d}z} = K\frac{2 M}{s} \sum_q
\biggl\{&[f_{q|P_1}(\xi_1,M)f_{\bar q|P_2}(\xi_2,M)
+ f_{\bar q|P_1}(\xi_1,M)f_{q|P_2}(\xi_2,M)]
\frac{{\rm d}\hat \sigma^\text{even}_{q \bar q}}{{\rm d}z} \nonumber \\
+&[f_{q|P_1}(\xi_1,M)f_{\bar q|P_2}(\xi_2,M)
- f_{\bar q|P_1}(\xi_1,M)f_{q|P_2}(\xi_2,M)]
\frac{{\rm d}\hat \sigma^\text{odd}_{q \bar q}}{{\rm d}z}\biggr\},
\nonumber \\
\end{aligned} \\
&\begin{aligned} \frac{{\rm d}\sigma_{gg}}{{\rm d}M\,{\rm d}y\,{\rm d}z} 
= K\frac{2 M}{s}\, f_{g|P_1}(\xi_1,M)f_{g|P_2}(\xi_2,M)
\frac{{\rm d}\hat\sigma_{gg}}{{\rm d}z}.
\end{aligned}
\end{align}
Here, ${\rm d}\hat\sigma_{q \bar q}^\text{even}/{\rm d}z$ and
${\rm d}\hat\sigma_{q\bar q}^\text{odd}/{\rm d}z$ are the even and
odd parts (under $z\leftrightarrow -z$) of the partonic
differential cross section ${\rm d}\hat\sigma_{q \bar q}/{\rm d}z$.
Furthermore, the $K$-factor accounts for higher order QCD corrections 
and, at NLO, can be approximated by the well-known expression
(see, for instance Ref.~\cite{Carena:2004xs})
\begin{equation}
K=1+\frac{4}{3}\,\frac{\alpha_s}{2\pi}\left(1+\frac{4}{3}\,\pi^2\right).
\label{Kfactor}
\end{equation}
For simplicity, and to make our procedure more transparent, 
we shall use in the sequel a global, flat, factor $K=1.3$.  
Although the full NLO corrections to the processes of interest here 
can require, as discussed in detail in Ref.~\cite{Mathews:2004xp}, 
a somewhat larger $K$-factor, especially for gluon-initiated processes, 
this effect would tend to cancel in the $A_\text{CE}$ asymmetry basic 
to our analysis, which is determined by ratios of angular-integrated 
(and mass-integrated around the resonance) cross sections. 
It may, however, have some bearing on the statistics, 
rendering the event rates based on the value  1.3 a slightly conservative 
estimate.
Finally,
$f_{j|P_i}(\xi_i,M)$ are parton distribution functions in the protons $P_1$
and $P_2$, and $\xi_i$ are the parton fractional momenta:
\begin{equation}
\xi_1=\frac{M}{\sqrt{s}}\, e^y, \qquad 
\xi_2=\frac{M}{\sqrt{s}}\, e^{-y}.
\end{equation}
In deriving Eq.~(\ref{Eq:dsigma-dMdydz}), the relations ${\rm d}\xi_1\,{\rm
d}\xi_2=(2M/s){\rm d}M {\rm d}y$ and $M^2=\xi_1\xi_2 s$ have been used, with
$s$ the $pp$ C.M.\ energy squared.  The minus sign in the odd term in that
equation allows us to interpret the angle $\theta_\text{cm}$ in the parton
cross section as being relative to the quark or gluon momentum (rather than
the proton momentum $P_1$).

\par The lepton differential angular distribution, for dilepton invariant mass
$M$ in an interval of size $\Delta M$ around the (narrow) resonance peak
$M_R$, is defined by
\begin{equation}
\frac{\dd\sigma}{\dd z}
=\int_{M_{R}-\Delta M/2}^{M_{R}+\Delta M/2}\dd M
\int_{-Y}^{Y}\frac{\dd\sigma}{\dd M\, \dd y\, \dd z}\,\dd y, 
\label{DiffCr}
\end{equation}
with $Y=\log(\sqrt{s}/M)$. 

\par The cross section for the narrow state production and subsequent decay
into a DY pair, $pp\to R \to l^+l^-$, is given by:
\begin{equation} 
\sigma(R_{ll})\equiv\sigma{(pp\to R)} \cdot
\text{BR}(R \to l^+l^-)
=\int_{-z_{\text{cut}}}^{z_\text{cut}}\dd z
\int_{M_{R}-\Delta M/2}^{M_{R}+\Delta M/2}\dd M
\int_{-Y}^{Y}\dd y
\frac{\dd\sigma}{\dd M\, \dd y\, \dd z}. \label{TotCr}
\end{equation}
Actually, if angular cuts are imposed by detector acceptance, $\vert
z\vert\leq z_{\rm cut}$, then $Y$ in Eqs.~(\ref{DiffCr}) and (\ref{TotCr})
must be replaced by some maximum value, $y_{\rm max}=y_{\rm max}(z)$.

\par One may notice that only terms in the partonic cross sections which are
even in $z$ contribute to the right-hand side of Eq.~(\ref{DiffCr}), because
in the case of the proton-proton collider odd terms do not contribute after
the integration over the rapidity $y$.  This holds true for the SM $\gamma
$-$Z$ interference term, as well as for SM-$G$ interference, with the SM
partonic cross section being pure $q\bar q$-initiated at the considered
order. Although such interference terms may appreciably contribute to the 
doubly-differential cross section $\dd\sigma/\dd M \dd z$, their contribution 
to the integral over $M$ needed in Eqs.~(\ref{DiffCr}) and (\ref{TotCr}),  
symmetrical around the graviton resonance mass $M_R$, 
is negligibly small for $M_Z\ll M_R$ and small resonance width 
(an approximation that will be assumed in the sequel), and negligible 
$M$-dependence of the overlap integral within the $\Delta M$ bin. 
This fact is pointed out for the case of $Z^\prime$s in, e.g., 
Refs.~\cite{Feldman:2006wb, Carena:2004xs}, but also holds for the graviton 
resonance case. 
Thus, in Eqs.~(\ref{DiffCr}) and (\ref{TotCr}), we can just retain the SM 
and the G pole contributions.

\par Keeping $z$-symmetric terms only, the partonic cross sections relevant to
the analysis presented below read \cite{Han:1998sg,Giudice:1998ck} (we follow
the notation of \cite{Dvergsnes:2004tw}): 
\begin{equation}
\frac{{\dd\hat{\sigma}_{q\bar q}^G }}{{\dd z}} 
+ \frac{{\dd\hat{\sigma}_{gg}^G }}{{\dd z}}\bigg\vert_{z{\ \rm even}} 
= \frac{{\kappa ^4 M^2 }}{{640\pi ^2 }}\left[ {\Delta _{q\bar q} (z) 
+ \Delta_{gg} (z)} \right]\vert\chi_G\vert^2,\label{partG} 
\end{equation}
\begin{equation}
\frac{\dd\hat \sigma_{q\bar q}^\text{SM}}{\dd z}\bigg\vert_{z{\ \rm even}}
= \frac{\pi \alpha_{\rm em}^2}{6 M^2}\,[S_q\, (1+z^{2})]. 
\label{partSM} 
\end{equation}
In Eq.~(\ref{partG}), $\chi_G$ represents the graviton $G$ propagator, with
$M_G$ and $\Gamma_G$ the mass and total width, respectively:
\begin{equation}
\chi_G=\frac{M^2}{{M^2 - M_G^2 + i\,M_G \Gamma _G }},
\label{propG}
\end{equation}
and, for the first massive mode, $\kappa$ is given by
\cite{Allanach:2000nr,Bijnens:2001gh,Dvergsnes:2002nc}
\begin{equation} \label{Eq:kappa-def}
\kappa  =  \sqrt 2\, \frac{{x_1 }}{{M_G }}\, c.
\end{equation}

\par The total width can be written as $ \Gamma_G=\rho\, x_1^2\, c^2\,M_G $,
where $\rho$ is a constant depending on the number of open decay
channels. Assuming the graviton decays only to the SM particles, and with
partial widths explicitly given in Refs.\
\cite{Allanach:2000nr,Han:1998sg,Bijnens:2001gh}, one finds
$\Gamma_G=1.43\,c^2\, M_G$. With $c\leq 0.1$ in the theoretically ``natural''
range, this value allows to use for the graviton resonance propagator the
narrow-width approximation,
\begin{equation}
\vert\chi_G\vert^2\rightarrow \delta(M-M_G)\,\frac{\pi
M_G^2}{2\,\Gamma_G}\label{NWA}.
\end{equation}

\par
The leading order angular dependencies in Eq.~(\ref{partG}) are given by
\begin{equation}
\Delta_{q\bar q} (z)
= \frac{\pi }{{8N_C }}\,\frac{5}{8}\,(1 - 3z^2  + 4z^4 ),
\hspace{1cm}
\Delta _{g g} (z) = \frac{\pi }{{2(N_C^2  - 1)}}\,\frac{5}{8}\,(1 - z^4),
\label{angfuncs}
\end{equation}
where $N_C$ is the number of quark colors.

\par
For the SM partonic cross section of Eq.~(\ref{partSM}) one has,
neglecting fermion masses:
\begin{equation} \label{Eq:SM-partonlevel}
S_q\equiv Q_q^2 Q_e^2 + 2 Q_q Q_e v_q v_e\, \Re\, \chi_Z
+(v_q^2+a_q^2)(v_e^2+a_e^2)\,|\chi_Z|^2,
\end{equation}
where, for fermion $f$, $a_f=T_{3f}$, $v_f=T_{3f} - 2Q_f \sin^2\theta_W$, and
the $Z$ propagator in the approximation $M\gg M_Z$ is represented by
\begin{equation}
\chi_Z (M)\approx \frac{1}{\sin^2(2 \theta_W)}\, \frac{M^2}{M^2 - M_Z^2}.
\end{equation}

\subsection{Statistical considerations}
\label{sect:statistical}
In the experimental discovery of a narrow resonance the observed width
is determined by the dilepton invariant mass resolution, that we 
may associate to the size of the bin $\Delta M$ introduced above. 
Clearly, on the one hand larger $\Delta M$  would allow a larger chance 
of detecting the resonance and, on the other hand, for a narrow resonance 
falling within the bin the integral over $\Delta M$ in Eq.~(\ref{TotCr}) 
should be practically insensitive from the size of $\Delta M$. 
Conversely, such an integral should be essentially proportional to the size of 
$\Delta M$ for the SM background.
This background is dominated by the SM Drell-Yan process, other SM
background contributions turn out to amount to at most a few percent of it
\cite{Cousins}.

\begin{table}[htb]
\begin{center}
\label{tab:identify}
\renewcommand{\tabcolsep}{.75em}
\caption{The number of signal events, $N_S$, in the RS model with
$c=0.01$ including ATLAS detector cuts as a function of resonance
mass $M_G$ in a run of 100 fb$^{-1}$ for the process $pp\to e^+e^-
+X$; the number of the SM background events, $N_B$, integrated
over the bin and the minimum number of signal events $N_S^\text{min}$
required to detect the resonance (at $5\sigma$).}
\vspace{.175in}
\begin{tabular}{|c|c|c|c|c|c|}
\hline
$M_G$ (GeV) & Bin $\Delta M$ (GeV) & $N_{S}$ & $N_{B}$& $N_S^\text{min}$ \\
\hline
1000 & 30.6  & 878.6  & 81.5 & 45.1 \\
1500 & 42.9 & 108.9 & 14.6 & 19.1 \\
1700 & 47.8 & 54.7 & 8.2 & 14.3 \\
1800 & 50.2 & 39.6 & 6.2 & 12.5 \\
1900 & 52.6 & 29.0 & 4.8 & 10.9 \\
2000 & 55.0 & 21.4 & 3.7 & 10.0 \\
2100 & 57.4 & 16.0 & 2.9 & 10.0 \\
2200 & 59.8 & 12.1 & 2.2 & 10.0 \\
2300 & 62.3 &  9.2 & 1.8 & 10.0 \\
2400 & 64.7 &  7.0 & 1.4 & 10.0 \\
2500 & 67.1 &  5.4 & 1.1 & 10.0 \\
\hline
\end{tabular}
\end{center}
\end{table}

\par Regarding the bin size, it depends on the energy resolution.  For the
ATLAS detector, the bin size $\Delta M$ at invariant dilepton mass $M$
measured in TeV units, can be parameterized as \cite{Atlas}:
\begin{equation} 
\Delta M = 24\, (0.625 M + M^2 +0.0056)^{1/2}\, {\rm GeV}.
\label{bin}
\end{equation}
For $M > 3$ TeV, the $M^2$ term dominates in Eq.~(\ref{bin}) and the bin size
grows linearly in $M$, so that ${\Delta M}\sim 24 M$ GeV for large
$M$. Similar results, comparable to about 10\%, hold for the CMS detector
\cite{CMS}. Throughout the paper we will use Eq.~(\ref{bin}) for the bin size.

\par At the LHC, with integrated luminosity ${\cal L}_{\rm
int}=100~{\rm{fb}}^{-1}$, the number of signal (resonant) events can be
computed by using $N_S = \sigma(R_{ll}) \,\epsilon_l\, {\cal L}_{\rm{int}}$
and the background events are defined as $N_B =N_{SM}$ (background integrated
over the bin). Here, $\epsilon_l$ is the experimental reconstruction
efficiency, taken to be 0.9 both for electrons and muons. To compute cross
sections we use the CTEQ6 parton distributions \cite{Pumplin:2002vw}. We
impose angular cuts relevant to the LHC detectors. The lepton pseudorapidity
cut is $|\eta|<\eta_{\rm{cut}}=2.5$ for both leptons (this leads to a
boost-dependent cut on $z$ \cite{Dvergsnes:2004tw}), and in addition to
the angular cuts, we impose on each lepton a transverse momentum cut
$p_\perp>p_\perp^{\rm{cut}}=20~{\rm GeV}$. 
Analogous to previous references, in the analysis given here,
we have adopted the criterion for the discovery limit that $5\sqrt{N_{B}}$ 
events or 10 events, whichever is larger, constitutes a signal. The
number of DY background events ($N_B$) inside each bin, the minimum number of
signal events required to detect a graviton resonance ($N_S^\text{min}$) and
the resonant signal events ($N_S$) at various $M_G$ are summarized in
Table~1. Only electron pairs are included in Table~1.

\begin{figure}[tbh] 
\centerline{ \hspace*{-6.0cm}
\includegraphics[width=17.cm,angle=0]{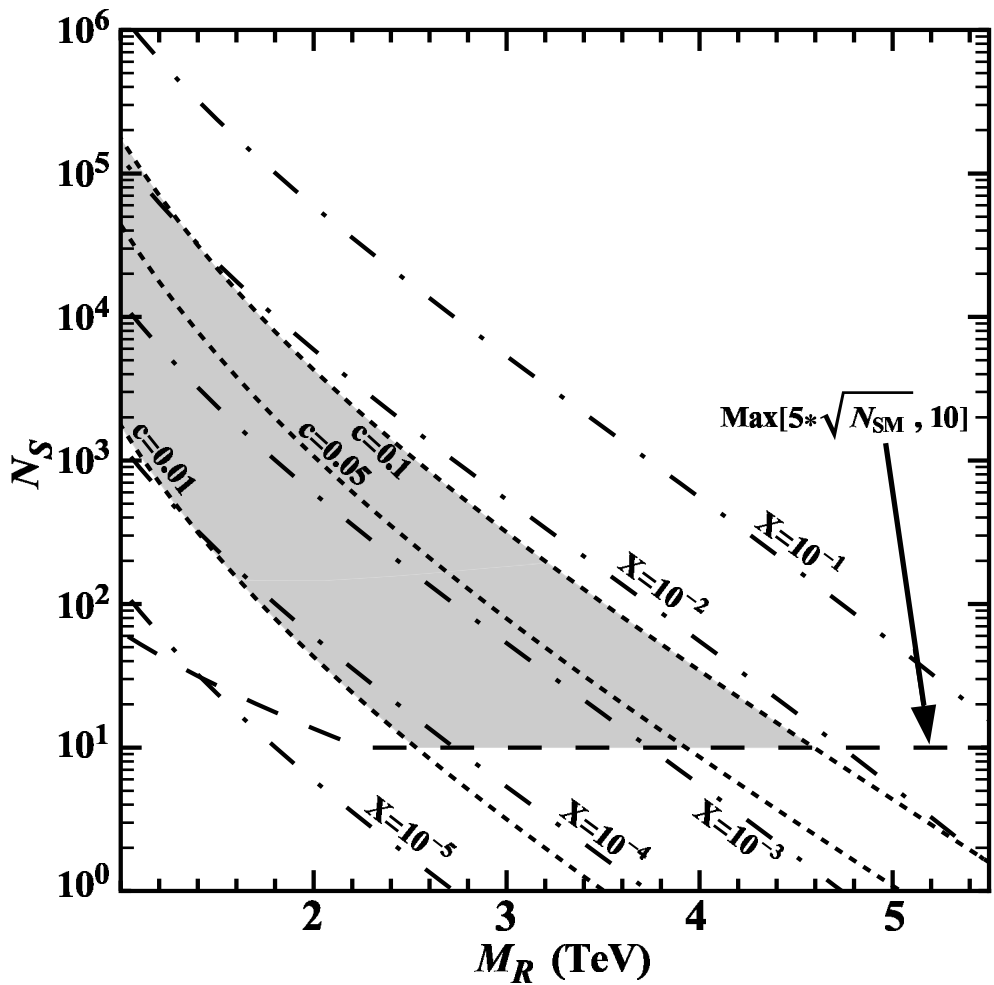}}
\vspace*{-6.3cm}
\caption{\label{Ns_vs_MG}
Expected number of
resonance (signal) events $N_S$ vs.\ $M_R$ ($R=G,\tilde\nu_\tau$)
at $\Lumint=100$ fb$^{-1}$ for graviton and sneutrino resonant
production with values of $c$= 0.01, 0.05, 0.1 (short dashed
curves) and $X$ (see Eq.~(\ref{Eq:X-def})) ranging from $10^{-5}$
to $10^{-1}$ in steps of 10 (dash-dotted curves) and the minimum number
of signal events (dashed curve) needed to detect the resonance
above the background in the process $pp\to l^+l^-+X$ ($l=e,\mu$).
Shaded area corresponds to potential overlap of graviton signature
space with that for sneutrino resonant production.}
\end{figure}

Fig.~\ref{Ns_vs_MG} shows the expected number of resonant (signal) events
$N_S$ {\it vs.}\ resonance mass $M_R$ ($R=G$) at ${\cal L}_{\rm{int}}=100$
fb$^{-1}$ for graviton production with values of $c$= 0.01, 0.05, 0.1 (dashed
curves), and the minimum number of signal events needed to detect it above the
background. With the assumption of efficiencies as stated above one finds
that, with 100 ${\rm{fb}}^{-1}$ of integrated luminosity, one can explore a
massive graviton up to a mass of about 2.5 TeV with $c=0.01$ ($5\sigma$
level), and this limit can be pushed to $\approx$ 4.5 TeV with $c=0.1$,
consistent with the results of \cite{Davoudiasl:2000jd}. 
While the analysis above is for the specific RS model, the general
features of this analysis may hold for a wider class of models which support
narrow resonances and predict spin-2 intermediate states.  We shall refer to a
region in the space spanned by resonance mass and number of events, that can
be populated by a certain model, as the ``signature space'' of that model. We
now proceed to sketch the competing (with the spin-2 resonance) non-standard
spin-0 and spin-1 interactions, and their respective signature spaces.

\section{Signature spaces of RS $G$ and sneutrino in $\rpv$}
\label{sect:sneutrino}
As mentioned in Sec.~\ref{sect:introduction}, models based on $\rpv$ SUSY can
mimic the RS graviton in a certain part of the parameter space as far as the
mass and narrowness of the resonance is concerned.  At tree-level, the
relevant parton process for DY lepton-pair production is in $R$-parity
breaking given by spin-0 sneutrino ($\tilde\nu$) formation from
quark-antiquark annihilation and subsequent leptonic decay:
\begin{equation} \label{Eq:qqbar-sneutrino}
q\bar{q} \to {\gamma,Z,\tilde{\nu}} \to l^+l^-.
\end{equation}
The corresponding partonic cross section is given by
\cite{Kalinowski:1997bc}
\begin{equation}
\frac{{\dd\hat{\sigma}_{q\bar q}}}{{\dd z}}= \frac{\dd\hat
\sigma_{q\bar q}^\text{SM}}{\dd z}
+ \frac{\dd\hat{\sigma}_{q\bar q}^{\tilde{\nu}}}{{\dd z}},
\label{part_snu}
\end{equation}
where the pure resonant term reads
\begin{equation}
\frac{\dd\hat{\sigma}_{q\bar q}^{\tilde{\nu}}}{{\dd z}}
= \frac{1}{3}\,\frac{\pi\alpha_\text{em}^2}{4\,M^2}
\left(\frac{\lambda\lambda'}{e^2}\right)^2
\vert\chi_{\tilde\nu}\vert^2\,\delta_{qd}. \label{cross_snu}
\end{equation}
Here, the propagator of the sneutrino $\chi_{\tilde\nu}$ is represented by
\begin{equation}
\chi_{\tilde\nu}=\frac{M^2}{{M^2 - M_{\tilde\nu}^2 +
i\,M_{\tilde\nu} \Gamma _{\tilde\nu} }}, \label{propNU}
\end{equation}
$M_{\tilde\nu}$ ($\Gamma _{\tilde\nu}$) is the mass (total decay width) of the
sneutrino, $\lambda'$ and $\lambda$ are the relevant $R$-parity-violating
couplings of $d\bar{d}$ and $l^+l^-$ to the sneutrino, respectively. We note
that the process (\ref{Eq:qqbar-sneutrino}), where the intermediate state is a
sneutrino, requires {\it two} $R$-parity-violating couplings to be
non-zero.\footnote{A different scenario was investigated in
\cite{Choudhury:2002av}, where only {\it one} such coupling was assumed
non-zero. Then a squark would be exchanged in the $t$- or $u$-channel, and the
angular distribution would be rather different.}  For the present case, the
$K$-factor has been studied for a range of sneutrino masses, and for different
parton distribution functions \cite{Choudhury:2002aua}. The value adopted for
the graviton case, $K=1.3$, remains a good approximation.

In the narrow width approximation the $\tilde\nu$-exchange cross section
(\ref{cross_snu}) can be written as:
\begin{equation}
\frac{\dd\hat{\sigma}_{q\bar q}^{\tilde{\nu}}}{{\dd z}}\approx
\frac{\pi}{24}\,\frac{X}{M_{\tilde\nu}}\,\delta(M-M_{\tilde\nu})\,\delta_{qd},
\label{cross_snu_NWA}
\end{equation}
where
\begin{equation} \label{Eq:X-def}
X=(\lambda^\prime)^2B_l.
\end{equation}
Here $B_l$ is the sneutrino leptonic branching ratio and $\lambda^\prime$ the
relevant coupling to the $d\bar{d}$ quarks. Indeed, due to $SU(2)$ invariance,
the sneutrino, which is a $T_3=+\half$-member of the doublet, can only couple
to a {\it down}-type quark. This model depends therefore on two independent
parameters, i.e., the sneutrino mass $M_{\tilde\nu}$ and $X$. With $i,j,k$
generation indices, the $R$-parity-violating coupling of interest is
$\lambda_{ijk}^\prime=\lambda_{i11}^\prime$, with $i$ denoting the sneutrino
generation. Among these, $\lambda_{111}^\prime$ is rather constrained, whereas
$\lambda_{211}^\prime$ and $\lambda_{311}^\prime$ could be as large as
$10^{-1}$--$10^{-2}$ for a 100~GeV sneutrino, and larger for a heavier one
\cite{constraints}.

\begin{figure}[htb] 
\centerline{
\includegraphics[width=15.0cm,angle=0]{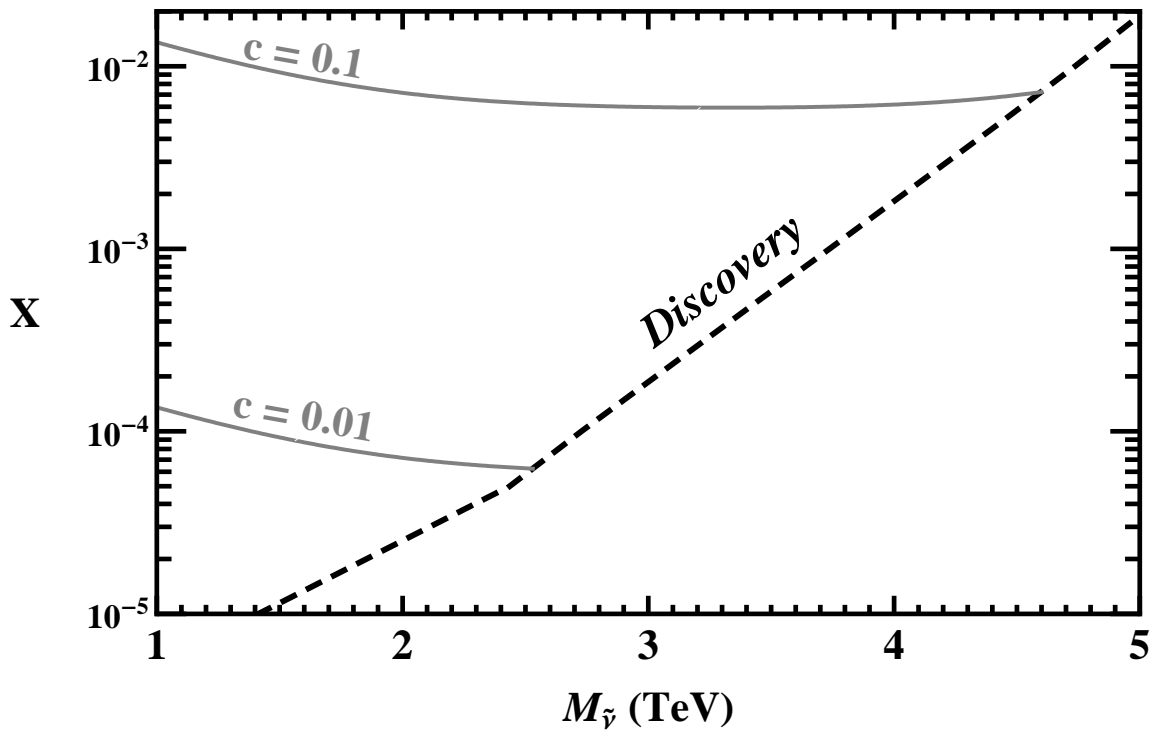}}
\caption{\label{X_vs_Msneu} The discovery reach at the $5\sigma$ level in the
plane ($M_{\tilde\nu}$,$X$) obtained from lepton pair production ($l = e,
\mu$) at the LHC with $\Lumint=100$ fb$^{-1}$. The discovery limit is defined
by $5\sqrt {N_{SM}}$ or by 10 events, whichever is larger. The kink in the
plot is the point of transition between the two criteria.  Indicated is the
domain in sneutrino parameters for discovery in the reach of LHC. The area
enclosed between the two solid lines and the dashed line corresponds to the
shaded area shown in Fig.~\ref{Ns_vs_MG}.}
\end{figure}
Quantitatively, the current constraints on $X$ are rather loose. The number of
signal events as a function of the spin-0 mass for the case of sneutrino
production with values of $X$ ranging from $10^{-5}$ to $10^{-1}$ in steps of
10 (dash-dotted curves), are given in Fig.~\ref{Ns_vs_MG}. The calculation has
been performed under the assumptions and kinematical cuts exposed in
Sec.~\ref{sect:statistical}. From Fig.~\ref{Ns_vs_MG} one can easily obtain
the discovery reach on sneutrino parameters ($5\sigma$ level) and translate
them into the plane ($M_{\tilde\nu}$,$X$) exhibited in
Fig.~\ref{X_vs_Msneu}. In this figure, the discovery region is on the left of
the dashed line, and the gray ($\sim$~horizontal) lines limit the ``confusion''
domain with the RS graviton in event rates, with $c=0.01$ and $c=0.1$,
respectively, see also Ref.~\cite{Allanach:1999bf}.

\par In Fig.~\ref{Ns_vs_MG}, the shaded area indicates the overlap of the
LHC-discovery parameter space for the $\rpv$ scenario {\it via}
$\sigma(pp\to\tilde{\nu} \to l^+ l^-)$ and that of the lowest RS graviton
scenario {\it via} $\sigma(pp \to G \to l^+ l^-)$. The figure indicates that,
as far as the total production cross section of DY dilepton pairs is
concerned, there exists a significantly extended domain in the
($M_{\tilde\nu},X$) plane where sneutrino $\tilde\nu$ production can mimic RS
graviton $G$ formation in its theoretically ``natural'' domain ($M_G,c$): in
these respective domains, the two scenarios can lead to the same number of
events under the resonance peak, $N_S(G\to l^+l^-)=N_S(\tilde{\nu}\to
l^+l^-)$. In other words, the two models are indistinguishable in
the overlapping domains of their parameter spaces, indicated by 
the shaded area in
Fig.~\ref{Ns_vs_MG}. Clearly, outside the ``common'' shaded area, the two
scenarios might be differentiated by means of event rates. For the
identification, the two models must be discriminated in the ``confusion''
region in Fig.~\ref{Ns_vs_MG}, this can be done by the spin determination of
the RS resonance.
\section{Signature spaces of RS $G$ and $Z^\prime$}
\label{sect:zprime}
Turning now to spin-1 resonance exchange, the differential cross section for
the relevant partonic process $q\bar{q} \to {\gamma,Z,Z^\prime} \to l^+l^-$
reads at leading order 
\begin{equation} 
\frac{{\dd\hat{\sigma}_{q\bar q}}}{{\dd z}}
= \frac{\dd\hat \sigma_{q\bar q}^\text{SM}}{\dd z} 
+ \frac{\dd\hat{\sigma}_{q\bar q}^{Z^\prime}}{{\dd z}},
\label{part_zprime}
\end{equation}
with
\begin{equation}
\frac{\dd\hat \sigma_{q\bar q}^{Z^\prime}}{\dd z}\bigg\vert_{z{\rm -even}}
= \frac{\pi \alpha_{\rm em}^2}{6 M^2}\,[S_q^\prime\, (1+z^{2})],
\label{partZprime}
\end{equation}
and, neglecting fermion masses:
\begin{equation} \label{Eq:zprime-partonlevel}
S_q^{Z^\prime}\equiv (v_q'^2+a_q'^2)(v_e'^2+a_e'^2)\,
\vert\chi_{Z^\prime}\vert^2.
\end{equation}
Here, we have introduced the $Z^\prime$ vector and axial-vector couplings to
SM fermions, and the $Z^\prime$ propagator $\chi_{Z^\prime}$ is represented by
\begin{equation}
\chi_{Z^\prime}=\frac{M^2}{{M^2 - M_{Z^\prime}^2 +
i\,M_{Z^\prime} \Gamma _{Z^\prime} }}. \label{prozprime}
\end{equation}
According to previous arguments, in the sequel we neglect
$(\gamma,Z)-Z^\prime$ interference terms in the cross section. Moreover,
effects from a potential $Z-Z^\prime$ mixing are also disregarded. 

\par As anticipated in Sec.~\ref{sect:introduction}, in addition to a generic
spin-1 exchange, we will consider the discrimination reach on the spin-2
lowest RS resonance from the, rather popular and physically motivated,
$Z^\prime$ scenarios where the couplings in Eq.~(\ref{Eq:zprime-partonlevel})
are constrained to have fixed values. One such model is the so-called
sequential model (SSM), where the $Z^\prime$ couplings to fermions are the
same as those of the SM $Z$.

\par Furthermore, we will consider: (i) the three possible $U(1) \, Z'$
scenarios originating from the exceptional group $E_{6}$ breaking; and (ii)
the $Z^\prime$ predicted by a left-right symmetric model that can originate
from an $SO(10)$ GUT. While detailed descriptions of these models can be
found, e. g., in Ref.~\cite{Hewett:1988xc}, we just recall that the three
heavy neutral gauge bosons are denoted by $Z^\prime_{\chi}$, $Z^\prime_{\psi}$
and $Z^\prime_{\eta}$ with specific coupling constants to SM matter displayed
in Table~\ref{Table:couplings}.
Regarding the case (ii), the mentioned left-right (LR)
model predicts a heavy neutral gauge boson $Z'_{LR}$ generally coupled to a
linear combination of the right-handed and $B$--$L$ currents [$B$ and $L$ are
baryon and lepton numbers, respectively]:
\begin{equation}
J_{LR}^\mu = \alpha_{LR} J^\mu_{3R} - (1/2\alpha_{LR}) J^\mu_{B-L}
\ \ {\rm with} \ \alpha_{LR}= \sqrt{ (c_W^2 g_R^2/s_W^2 g_L^2)-1}.
\end{equation}
Here, $g_L$=$e/s_W$ and $g_R$ are the ${\rm SU(2)_L}$ and ${\rm SU(2)_R}$
coupling constants with $s_W^2= 1-c_W^2 \equiv \sin^2\theta_W$; the parameter
$\alpha_{LR}$ is restricted to the range $\sqrt{2/3} \lsim \alpha_{LR} \lsim
\sqrt{2}$. The upper bound corresponds to the so-called LR-symmetric
$Z^\prime_{\rm LR}$ model with $g'_R=g'_L$, while the lower bound is found to
coincide with the $Z^\prime_\chi$ model introduced above.

\par Finally, we will include in our analysis the case of the $Z'_{\rm ALR}$
predicted by the so-called ``alternative'' left-right scenario
\cite{Hewett:1988xc,Cvetic:1995zs}.

\par All numerical values of the $Z^\prime$ couplings needed in
Eq.~(\ref{Eq:zprime-partonlevel}) are collected in
Table~\ref{Table:couplings}, where: $v^\prime_f
(a^\prime_f)=(g_L^{f\prime}\pm g_R^{f\prime})/2$; $A=\cos\beta/2\sqrt 6$,
$B=\sqrt {10}\sin\beta/12$ with $\beta =0$, $\pi/2$ and $\arctan(-\sqrt{5/3})$
for the $Z^\prime_{\chi}$, $Z^\prime_{\psi}$ and $Z^\prime_{\eta}$,
respectively. We have introduced the notations ${g_Z}^\prime=1/c_W$ for the
$E_6$ and the LR models and ${g_Z}^\prime=1/({s_W\, c_W\sqrt{1-2s_W^2})}$ for
the ALR model.  Current direct search limits on $Z^\prime$ masses from the
Fermilab Tevatron are of the order of 900~GeV or less~\cite{Tev:2007sb}.

\begin{table}[htb]
\caption{\label{Table:couplings} Left-handed and right-handed couplings of the
first generation of SM fermions to the $Z^\prime$ gauge bosons, needed in
Eq.~(\ref{Eq:zprime-partonlevel}).}
\begin{center}
\begin{tabular}{|c|c|c|c|c|} \hline
\multicolumn{5}{|c|}{$E_6$ model}\\ \hline \hline
fermions ($f$)  & $\nu$ & $e$ & $u$ & $d$ \\
\hline 
$g_L^{f\prime}/{g_Z}^\prime$ & $3A+B$ & $3A+B$ & $-A+B$ & $-A+B$ \\
\hline
$g_R^{f\prime}/{g_Z}^\prime$ & 0 & $A-B$ & $A-B$ & $-3A-B$ \\
\hline \hline \multicolumn{5}{|c|}{Left-Right model (LR)}\\
\hline \hline $g_L^{f\prime}/{g_Z}^\prime$ &
${\frac{1}{2\,\alpha_{LR}}}$ & 
${\frac{1}{2\,\alpha_{LR}}}$ & 
$-{\frac{1}{ 6\,\alpha_{LR}}}$ & 
$-{\frac{1}{6\,\alpha_{LR}}}$ \\
\hline $g_R^{f\prime}/{g_Z}^\prime$ & 0 & 
${\frac{1}{2\,\alpha_{LR}}}- {\frac{\alpha_{LR}}{ 2}}$ & 
$-{\frac{1}{6\,\alpha_{LR}}}+{\frac{\alpha_{LR}}{2}}$ & 
$-{\frac{1}{6\,\alpha_{LR}}}-{\frac{\alpha_{LR}}{2}}$ \\
\hline \hline \multicolumn{5}{|c|}{Alternative Left-Right model (ALR)}\\
\hline \hline $g_L^{f\prime}/{g_Z}^\prime$ &
${-\frac{1}{2}+s_W^2}$ & ${-\frac{1}{2}+s_W^2}$ & 
${-\frac{1}{6}s_W^2}$ &
${-\frac{1}{6}s_W^2}$ \\
\hline
$g_R^{f\prime}/{g_Z}^\prime$ & 0 & 
${-\frac{1}{2}+\frac{3}{2}s_W^2}$ & 
${\frac{1}{ 2}-\frac{7}{6}s_W^2}$ & $\frac{1}{3}s_W^2$ \\
\hline
\end{tabular}
\end{center}
\end{table}

\par The $Z'$ partial decay widths into massless fermion-antifermion pairs in
$E_6$ and LR models are functions of $\beta$ and $\alpha_{LR}$,
respectively. From the analysis of Refs.~\cite{Hewett:1988xc,Pankov:1992cy} it
turns out that, in the absence of ``exotic'' decay channels, the total width
$\Gamma_{Z^\prime}\ll M_{Z^\prime}$, so that the narrow width approximation to
the $Z^\prime$ propagator should be adequate for our numerical estimates.

\begin{figure}[tbh] 
\centerline{ \hspace*{.0cm}
\includegraphics[width=12.0cm,angle=0]{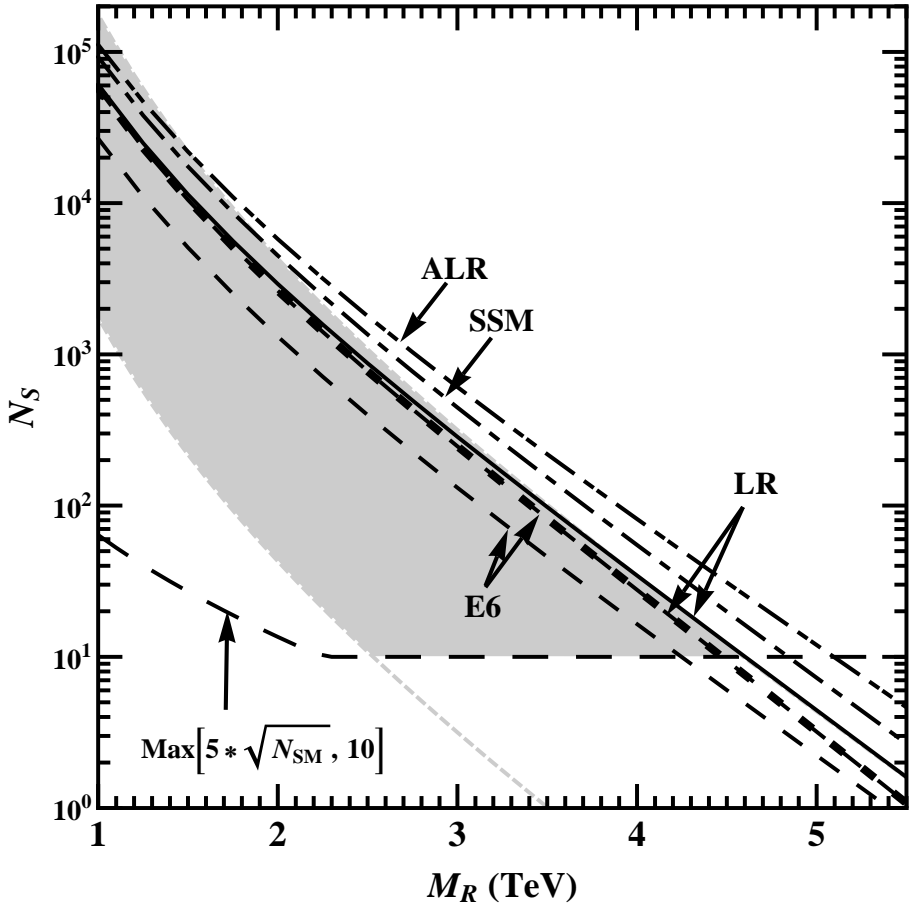}}
\vspace*{.0cm} 
\caption{\label{figZprime} Same as in Fig.~\ref{Ns_vs_MG} but for number of
resonance events $N_S$ vs.\ $M_R$ ($R=G,Z^\prime$) for graviton and $Z^\prime$
resonant production and the minimum number of signal events needed to detect
the resonances above the background in the process $pp\to l^+l^-+X$
($l=e,\mu$). The two ``LR'' lines refer to the extreme values for
$\alpha_{LR}$.  The shaded area is the overlap of graviton and sneutrino
signature spaces for $0.01<c<0.1$, with ${\cal L}_{\rm int}=100$ fb$^{-1}$.}
\end{figure}
\par
\begin{figure}[htb] 
\centerline{ \hspace*{0.0cm}
\includegraphics[width=12.0cm,angle=0]{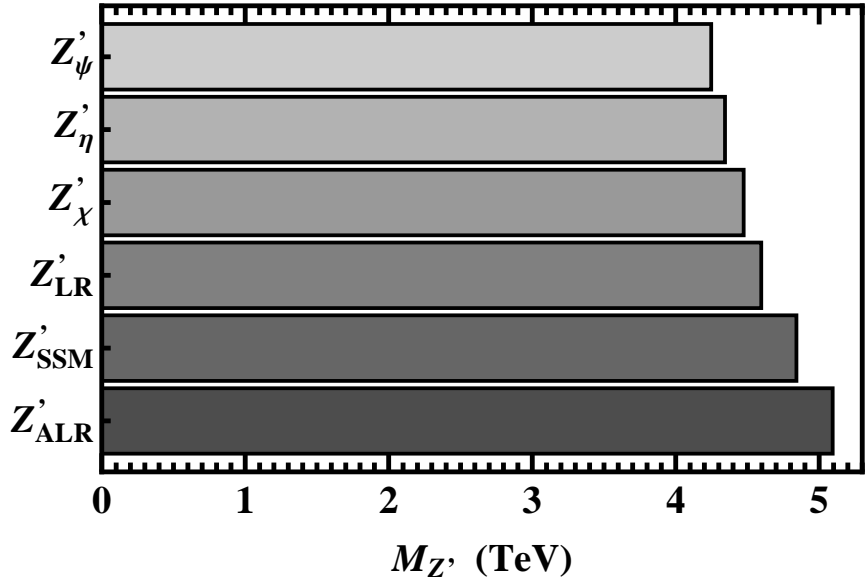}}
\vspace*{0.0cm} 
\caption{\label{HistogrZprime} Discovery limits at the $5\sigma$ level for
neutral gauge bosons of representative models, obtained from lepton pair
production ($l = e, \mu$) at the LHC with $\Lumint=100$ fb$^{-1}$.}
\end{figure}

 
The number of $Z'$ signal events as a function of resonance mass for the
representative models summarized in Table~\ref{Table:couplings}, and LHC
luminosity of 100 fb$^{-1}$, are given in Fig.~\ref{figZprime}. From this
figure, one can easily obtain the $5\sigma$ level discovery reaches on the
corresponding $Z^\prime$ masses, presented as a histogram in
Fig.~\ref{HistogrZprime}. These estimates are numerically consistent with
those in Refs.~\cite{Feldman:2006wb,Hewett:1988xc} and
\cite{Weiglein:2004hn,Cvetic:1995zs,Godfrey:2002tna,Dittmar:2003ir,
Rizzo:2006nw,Petriello:2008zr}. 

\par In these cases, the $Z^\prime$ signature spaces reduce to lines, and
Fig.~\ref{figZprime} shows that, at the assumed LHC luminosity, the lowest, RS
spin-2, resonance can be discriminated against the ALR and SSM spin-1
$Z^\prime$ scenarios already at the level of event rates in a large range of
$M_{Z^\prime}$ values, with no need for further analyses based on angular
distributions. Only the $E_6$ and LR $Z^\prime$ models possess a ``confusion
region'' with the RS resonance $G$, concentrated near the upper border of the
graviton allowed signature domain. This may represent an interesting
information by itself.

\par In the next sections, we turn to the identification of the spin-2 of the
first RS resonance, {\it vs.} the spin-1 and spin-0 hypotheses.

\section{Angular distributions in the dilepton channel}
\label{sect:angdist}
The normalized angular distributions of the relevant parton processes,
mediated by spin-2, spin-1 and spin-0 formation and subsequent decay to DY
pairs, are shown in Table~\ref{tab:AngDist}, as summarized, e.g., also in
Refs.~\cite{Allanach:2002gn,Cousins:2005pq}.

\begin{table}[tbh]
\begin{center}
\caption{\label{tab:AngDist}
Normalized angular distributions for the decay products
of spin-0, spin-1 and spin-2 resonances, considering only even
terms in $z\equiv\cos\theta_\text{cm}$ for parton subprocesses.}
\vspace{.175in}
\renewcommand{\tabcolsep}{.75em}
\begin{tabular}{|c|c|c|}
\hline Process  & Normalized density for $\cos\theta_\text{cm}$,
$(1/\hat\sigma)\dd\hat\sigma/\dd z$  \\  \hline
$q\bar{q}\to(\gamma, Z) \to l^+l^-$
& \phantom{$\Bigg|$}
${\displaystyle{\frac{3}{8}\,(1+z^2)}}$  \\
$q\bar{q}\to Z^\prime\to l^+l^-$ &   \\
\hline
$gg\to G\to l^+l^-$
&\phantom{$\Bigg|$}
${\displaystyle{\frac{5}{8}\,(1-z^4)}}$  \\
\hline $q\bar{q}\to G\to l^+l^-$ &
\phantom{$\Bigg|$}
${\displaystyle{\frac{5}{8}\,(1-3\,z^2+4\, z^4)}}$  \\
\hline $\bar q q\to{\tilde{\nu}}\to l^+l^-$ & \phantom{$\Bigg|$}
${\displaystyle{\frac{1}{2}}}$  ({\rm flat}) \\
\hline
\end{tabular}
\end{center}
\end{table}

For simplicity, and according to the considerations made in
Sec.~\ref{sect:crossstatistics}, in this table only the $z$-even terms in the
parton differential cross sections are retained, $z$-odd contributions
disappear from the observables we will consider.

\par
The correspondence between spin and angular distribution is quite sharp: a
spin-0 resonance determines a flat angular distribution, spin-1 corresponds to
a parabolic shape, and spin-2 yields a quartic distribution. The CDF
collaboration has recently attempted angular distribution analyses using the
cumulative DY data at the $p\bar p$ Tevatron collider, their results are
reported in Ref.~\cite{Abulencia:2005nf}. The LHC promises tests of the spin
hypotheses with significantly higher sensitivity, due to the definitely higher
statistics allowed by the foreseen larger energy and luminosity.

\begin{figure}[htb] 
\centerline{ \hspace*{-2.5cm}
\includegraphics[width=13.0cm,angle=0]{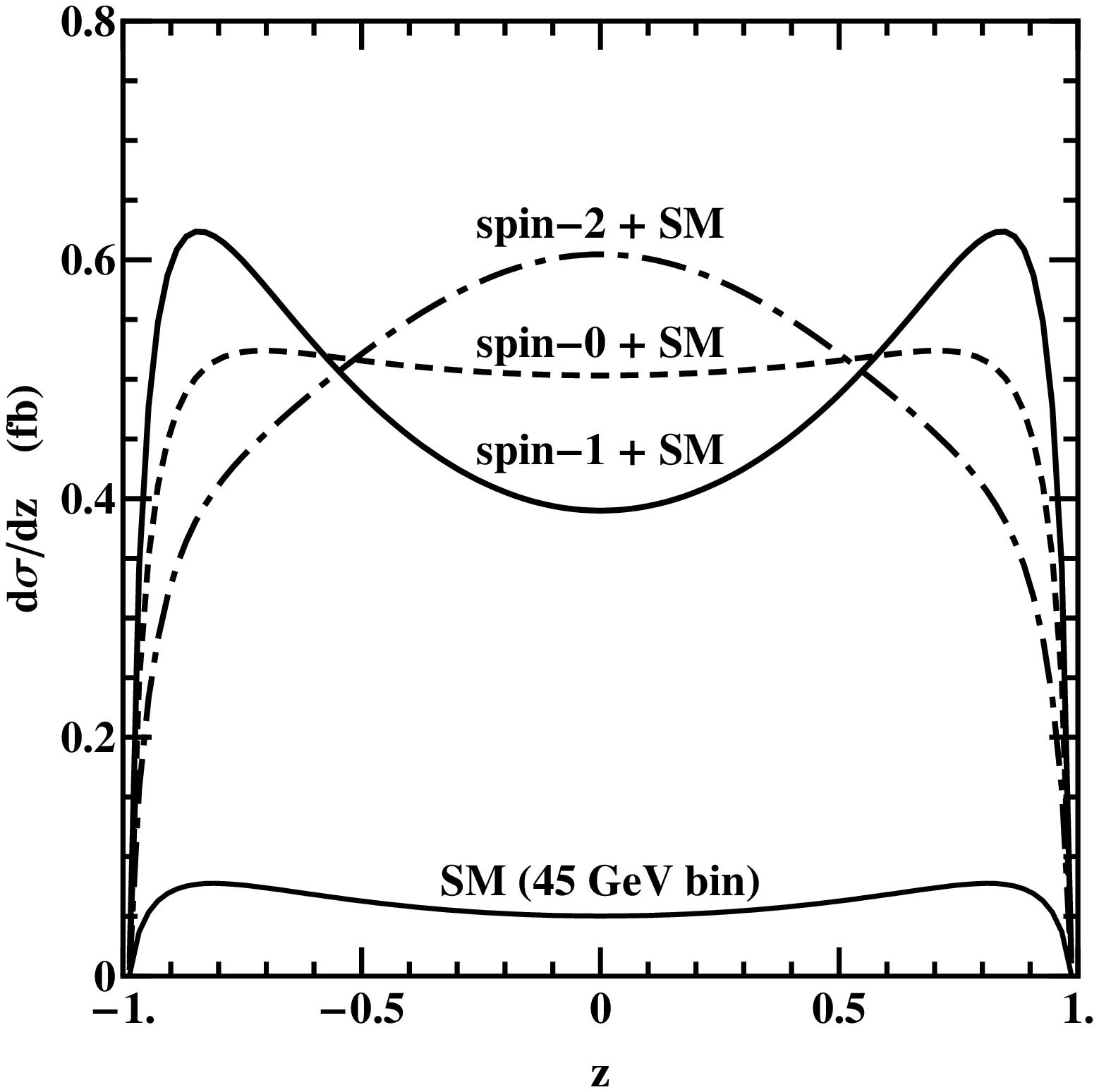}}
\vspace*{-0.cm}
\caption{\label{Ang_dist}
Angular distribution of leptons in the dilepton center of mass system 
for (i) spin-2 graviton resonant production,
$\dd \sigma(pp \to G \to l^+ l^-)/ \dd z$, in the RS model
with $c$=0.01; (ii) spin-0 
resonant production, $\dd \sigma(pp \to 
S \to l^+ l^-)/ \dd z$;
(iii) spin-1 
resonant production, $\dd \sigma(pp 
\to V \to l^+ l^-)/ \dd z$. We take $M_R=1.6$ TeV
and assume equal numbers of resonant DY events.}
\end{figure}

Using Eq.~(\ref{DiffCr}), one can derive the angular distributions determined
by spin-2 RS graviton resonance, spin-1 $V$ and spin-0 $S$,
respectively. These distributions can be conveniently written in a
self-explanatory way as [$G$ denotes the spin-2 resonance, while $V$ and $S$
denote the spin-1 and spin-0 cases, respectively]:
\begin{equation}
\label{Eq:AngulDistG}
\frac{\dd \sigma(G_{ll})}{\dd z}
= \frac{3}{8}\,(1 + z^2)\,\sigma_{q\bar q}^\text{SM}
+ \frac{5}{8}\,(1 - 3z^2 + 4z^4 )\,\sigma _{q\bar q}^G
+ \frac{5}{8}\,(1 - z^4)\, \sigma _{gg}^G,
\end{equation}
\begin{equation}\label{Eq:AngulDistZprime}
\frac{\dd{\sigma(V_{ll})}}{\dd z}
= \frac{3}{8}\,(1 + z^2)\,(\sigma_{q\bar q}^\text{SM}
+ \sigma_{q\bar q}^{V} ),
\end{equation}
\begin{equation}\label{Eq:AngulDistSNU}
\frac{\dd{\sigma(S_{ll})}}{\dd z}
= \frac{3}{8}\,(1 + z^2)\,\sigma_{q\bar q}^\text{SM}
+ \frac{1}{2}\, {\sigma}_{q\bar q}^{S}.
\end{equation}

\par Corresponding to Eqs.~(\ref{Eq:AngulDistG})--(\ref{Eq:AngulDistSNU}), the
integrated $pp$ production cross sections for the $G$, 
$V$ and $S$ hypotheses are given by
\begin{equation} \label{Eq:ProdCrSect}
\sigma(G_{ll}) =\sigma_{q\bar q}^\text{SM}
+\sigma _{q\bar q}^G + \sigma _{gg}^G, \quad
\sigma(V_{ll})
= \sigma_{q\bar q}^\text{SM} + \sigma_{q\bar q}^{V}, \quad
\sigma(S_{ll})
=\sigma_{q\bar q}^\text{SM} + {\sigma}_{q\bar q}^S.
\end{equation}

\par Detector cuts are not taken into account in the above
Eqs.~(\ref{Eq:AngulDistG})--(\ref{Eq:ProdCrSect}). We shall use these
relations for illustration purposes, in order to better expose the most
important features of the method we use. The final numerical results, as well
as the relevant figures that will be presented in the sequel refer to the full
calculation, with detector cuts taken into account. It turns out, however,
that such results are numerically close to those derived from the application
of Eqs.~(\ref{Eq:AngulDistG})--(\ref{Eq:ProdCrSect}).

\par The angular distributions arising from the spin-2, spin-1 and spin-0
resonances are represented in Fig.~\ref{Ang_dist}, for the same peak masses
$M_R$ in the three hypotheses and the same number of signal events, $N_S$,
under the peak.  The angular distributions in this figure are somewhat
distorted compared to those in Table~\ref{tab:AngDist}, because of (i)~the
smearing due to the parton distributions in the protons, (ii)~different
partons contribute with different weight to the different channels, and
(iii)~detector cuts are taken into account.
\section{Identification of the spin-2 of the RS graviton} 
\label{sect:gravitonidentification}

\subsection{Center-edge asymmetry} 
\label{sect:ace}
To assess the identification power of the LHC of distinguishing the spin-2 RS
resonance from both spin-1 and spin-0 exchanges, we adopt the integrated
center-edge asymmetry $A_{\rm CE}$ introduced in
Refs.~\cite{Dvergsnes:2004tw,Osland:2003fn}. Basically, the advantage of this
observable lies in its insensitivity to spin-1 exchanges in the
$s$-channel. This property follows from the fact that such exchanges are
characterized by the same $z$-distributions as the SM $\gamma$- and
$Z$-exchanges, see Eq.~(\ref{Eq:AngulDistZprime}). Thus, deviations of $A_{\rm
CE}$ from the SM predictions could be attributed to graviton exchanges and,
accordingly, one could expect a particularly high sensitivity in the
identification of this kind of effects. Also, being ``normalized'' to the
cross section integrated over angles, one may hope this observable to be less
sensitive to systematic uncertainties.

\par 
In the present application, we define the center-edge asymmetry, with
$R$ labelling the three hypotheses we want to compare, as:
\begin{equation}
\label{Eq:ace} A_{\rm{CE}}(M_R)=\frac{\sigma_{\rm{CE}}(R_{ll})}
{\sigma(R_{ll})},
\end{equation}
with the ``center minus edge'' cross section:
\begin{equation}
\label{Eq:sigma-ce} \sigma_{\rm{CE}} \equiv
\left[\int_{-z^*}^{z^*} - \left(\int_{-z_\text{cut}}^{-z^*}
+\int_{z^*}^{z_\text{cut}}\right)\right] \frac{\dd
\sigma(R_{ll})}{\dd z}\, \dd z.
\end{equation}
Here: $0<z^*<{z_\text{cut}}$ is a, a priori free, value of $\cos\theta_{\rm
cm}$ that defines the separation between the ``center'' and the ``edge''
angular regions; in the approximation $z_{\rm cut}=1$, ${\dd
\sigma(R_{ll})}/{\dd z}$ are given by
Eqs.~(\ref{Eq:AngulDistG})--(\ref{Eq:AngulDistSNU}) and the total cross
sections $\sigma(R_{ll})$ by Eq.~(\ref{Eq:ProdCrSect}).

\par We assume that a deviation from the SM is discovered in the cross section
for dilepton production at LHC in the form of a narrow peak in the dilepton
invariant mass, and attempt the determination of the domain in the RS
parameter space where such a peak can be {\it identified} as being caused by
the spin-2 RS exchange, and the spin-0 and spin-1 hypotheses excluded. We also
assume the integrated center-edge asymmetry evaluated within the RS model to
be consistent with the measured data, and call this spin-2 model the ``true''
or ``best-fit''model.  We want to assess the level at which this ``true''
model is distinguishable from the other hypotheses, with spin-0 and spin-1,
that can compete with it as sources of a resonance peak in dilepton production
yielding in particular the same number of signal events.

\par
\begin{figure}[bht] %
\centerline{ \hspace*{-1.0cm}
\includegraphics[width=7.5cm,angle=0]{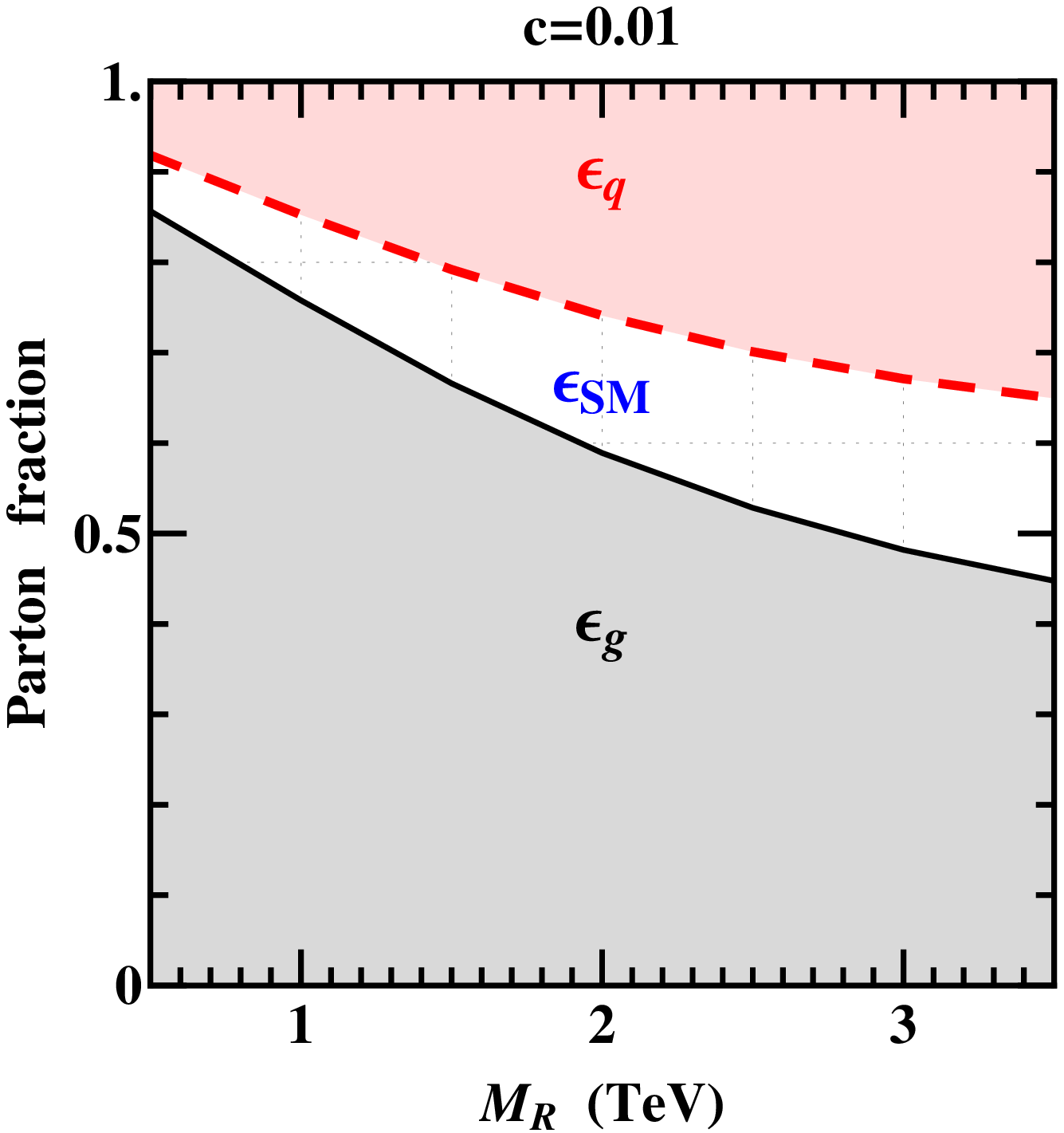}
\includegraphics[width=7.5cm,angle=0]{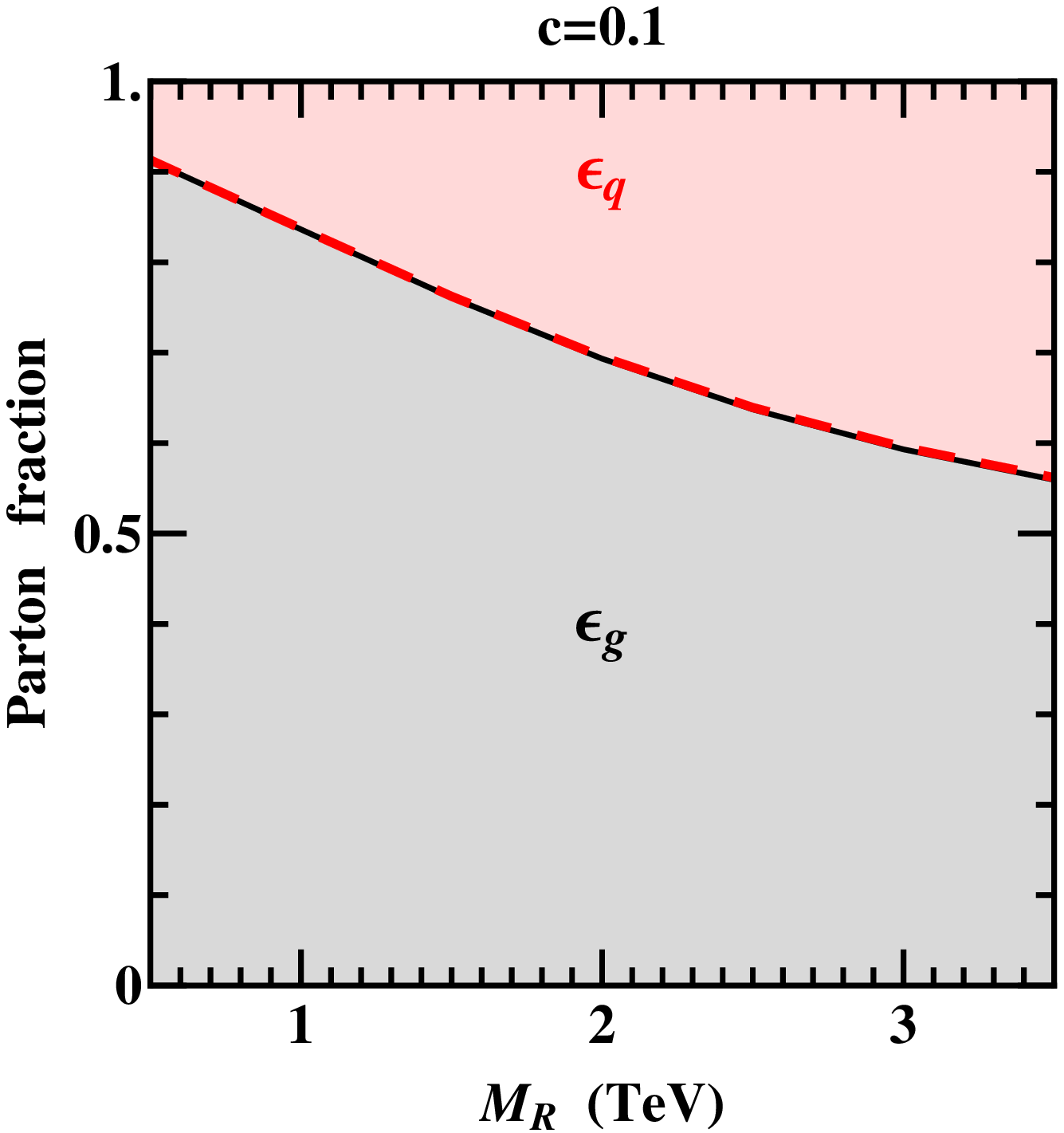}}
\vspace*{0.0cm} 
\caption{\label{epsilongg}  Contribution of gluon-gluon 
fusion $\epsilon_g^G$
and quark-antiquark annihilation $\epsilon_q^G$
to graviton production at the LHC as a function of mass,
displayed cumulatively, for $c=0.01$ and 0.1.}
\end{figure}

 
The explicit $z^*$-dependence of the center-edge asymmetries for the
three cases of interest here, obtained from
Eqs.~(\ref{Eq:AngulDistG})--(\ref{Eq:ProdCrSect}) and
Eqs.~(\ref{Eq:ace})--(\ref{Eq:sigma-ce}) are, in the same notations:
\begin{equation}\label{ACE2}
A_{\rm CE}^{G}
=\epsilon_q^{\rm SM}\,A_{\rm CE}^{V}
+ \epsilon_q^G\left[2\,{z^*}^5+\frac{5}{2}\,z^*(1-{z^*}^2)-1\right]
+ \epsilon_g^G\left[\frac{1}{2}\,{z^*}(5-{z^*}^4)-1\right],
\end{equation}
\begin{equation}\label{ACE1}
A_{\rm CE}^{V} \equiv A_{\rm CE}^{\text{\rm SM}}
=\frac{1}{2}\,z^*({z^*}^2+3)-1,
\end{equation}
\begin{equation}\label{ACE0}
A_{\rm CE}^{S}
= \epsilon_q^{\rm SM}\,A_{\rm CE}^{V}
+\epsilon_q^{S}\,(2\,z^*-1).
\end{equation}
Here, $\epsilon_q^G$, $\epsilon_g^G$s and $\epsilon_q^{\rm SM}$ are the
fractions of resonant events for $q\bar q,gg\to G\to l^+l^-$ and SM
background, respectively, with $\epsilon_q^G+\epsilon_g^G+ \epsilon_q^{\rm
SM}=1$. They are determined by the ratios of $\sigma_{q\bar q}^G$, etc., of
Eq.~(\ref{Eq:AngulDistG}) and $\sigma(G_{ll})$ of Eq.~(\ref{Eq:ProdCrSect}),
and shown in Fig.~\ref{epsilongg} for two values of $c$.

Analogous definitions hold for the other cases.  One should emphasize again
that, for spin-1, $A_{\rm CE}^{V}$ and the SM background (predominantly from
the DY continuum \cite{Cousins:2005pq}) have the same form, independent of
couplings and resonance mass.

\begin{figure}[htb] 
\centerline{ \hspace*{-0.3cm}
\includegraphics[width=7.8cm,angle=0]{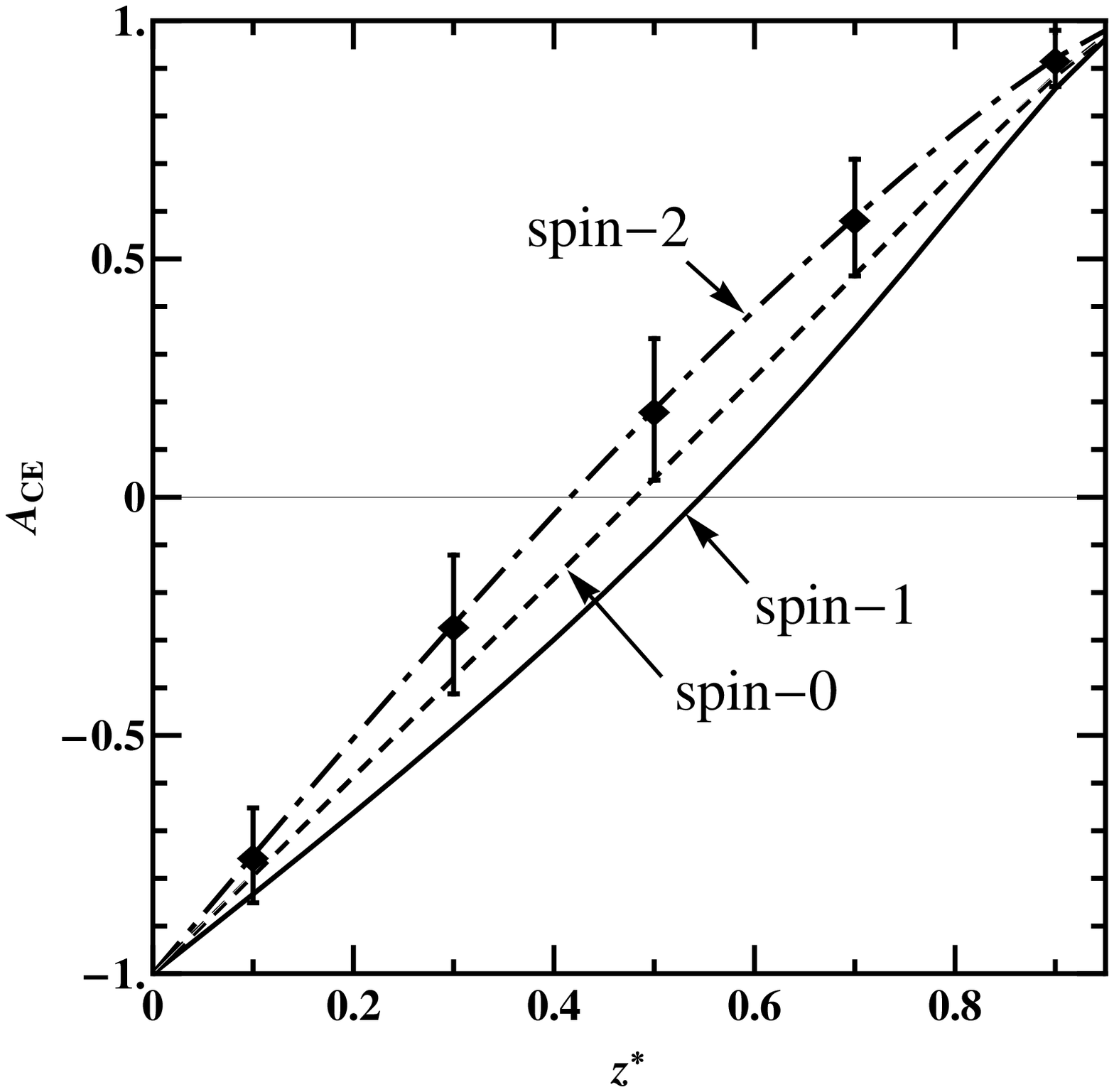}
\hspace*{0.0cm}
\includegraphics[width=7.8cm,angle=0]{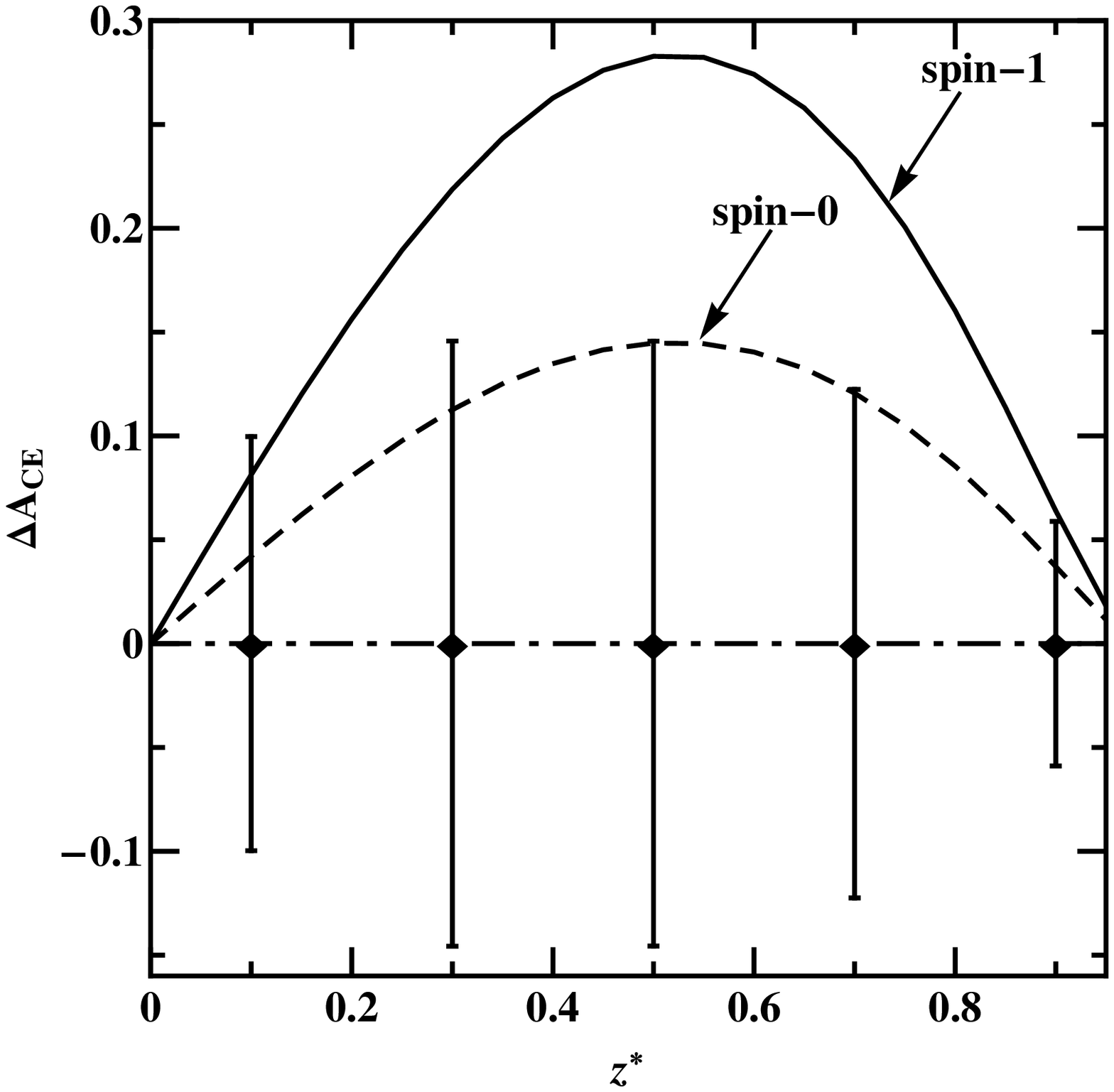}}
\vspace*{0.0cm} 
\caption{\label{ACE_vs_z} Left panel: $A_{\rm {CE}}$ {\it vs.}  $z^{*}$ for
the spin-2 resonance $G$ at $c$ =0.01 and $M_R$=1.6 TeV (dot-dashed curve),
and for the spin-0 (dashed curve) and spin-1 (solid curve) hypotheses, all for
the same $M_R$ and number of events. The error bar at $z^*=0.5$ is within the
identification reach on $G$ (at the $2\sigma$ level) for ${\cal L}_{\rm
int}=100$ fb$^{-1}$ as explained in the text.  Right panel: Asymmetry
deviations, $\Delta A_{\rm {CE}}$, of the spin-1 and spin-0 hypotheses from
the RS one, compared with the uncertainties on $A_{\rm {CE}}^{G}$.}
\end{figure}

\par As an example, in Fig.~\ref{ACE_vs_z} (left panel) the center-edge
asymmetry $A_{\rm CE}$ is depicted as a function of $z^*$ for resonances with
different spins, same mass $M_R=1.6$~TeV and same number $N_S$ of signal
events under the peak.  The dot-dashed curve corresponds to the spin-2 RS
graviton with $c=0.01$. The calculation is performed using the parton
distributions mentioned in Sec.~\ref{sect:statistical}, and detector cuts as
well as the SM background have been accounted for. Actually, the
$z^*$-behavior of $A_{\rm CE}(z^*)$ resulting from the full calculation is
found essentially equivalent to those presented in
Eqs.~(\ref{ACE2})--(\ref{ACE0}). Differences are appreciable only for $z^*$
close to 1, and turn out to have negligible impact on the numerical
determinations of the identification reaches presented in the sequel, where
the relevant chosen values of $z^*$ are in a range around 0.5. Indeed, since
numerically the $\chi^2$ turns out to have a smooth dependence there, for
definiteness we will present the results obtained from $A_{\rm CE}(z^*=0.5)$.

\par The deviations of the $A_{\rm CE}$ asymmetry from the prediction of the
RS model, caused by the spin-0 exchange
\begin{equation}  \label{Eq:Delta-ACE2-0}
\Delta
A_\text{CE}=A_\text{CE}^G-A_\text{CE}^S
\end{equation}
and that caused by the spin-1 exchange
\begin{equation}  \label{Eq:Delta-ACE2-1}
\Delta
A_\text{CE}=A_\text{CE}^G-A_\text{CE}^V,
\end{equation}
respectively, are depicted in Fig.~\ref{ACE_vs_z} (right panel). The
identification potential depends, of course, from the available statistics (as
well as on systematic uncertainties). In the example of Fig.~\ref{ACE_vs_z},
the vertical bars attached to the dot-dashed curve represent the 2$\sigma$
statistical uncertainty on the $A_{\rm CE}$ of the RS graviton model, assumed
to be the ``true'' model consistent with the data as stated above, with the
values of $M_G$ and $c$ reported in the caption and integrated LHC luminosity
of 100~fb$^{-1}$. One reads from Fig.~\ref{ACE_vs_z} that, at such (high)
luminosity, the spin-2 RS graviton with mass $M_G=1.6$~TeV and coupling
$c=0.01$ can, indeed, be discriminated from the other spin-hypotheses by means
of $A_\text{CE}$ at $z^*\simeq 0.5$.

\par Actually, Eqs.~(\ref{ACE1}) and (\ref{ACE0}) show the peculiar feature of
$A_{\rm {CE}}(z^*)$, that for same number of signal events:
\begin{equation}
A_{\rm CE}^S(z^*)>A_{\rm CE}^V(z^*),
\label{DeltaA1-0} 
\end{equation}
for all values of $0<z^*<1$.  This property is of course reproduced in
Fig.~\ref{ACE_vs_z}, and allows to conclude that, in order to identify the
spin-2 graviton resonance, if one is able to exclude the spin-0 hypothesis,
the whole class of spin-1 models will then automatically be excluded, so that
the spin-2 {\it identification} from the spin-1 hypothesis would be
model-independent. Stated in a statistical language, Eq.~(\ref{DeltaA1-0})
explicitly realizes the statement that discrimination of the spin-2 RS
resonance from the spin-1 hypothesis requires, for a given confidence level,
less events than the discrimination from the spin-0 one, as also noted in
Ref.~\cite{Cousins:2005pq}.

\subsection{Numerical results for RS graviton identification}
\label{sect:RSidentification}
We now consider the determination of the spin-2 of the resonance, based on the
assessment of the corresponding required minimal numbers of signal events
under the peak, $N_{\rm min}$.  To this purpose, we consider the deviations of
the (assumed to have been measured) center-edge asymmetry $A_\text{CE}^{G}$ from
those expected from pure spin-0 exchange, $A_\text{CE}^{S}$, and from spin-1
exchange, $A_\text{CE}^{V}$, defined by Eqs.~(\ref{Eq:Delta-ACE2-0}) and
(\ref{Eq:Delta-ACE2-1}), respectively.

\par Eqs.~(\ref{ACE2})--(\ref{ACE0}) continue being a useful representation of
these matters and accordingly, before presenting results from full
calculations, we write the deviation of Eq.~(\ref{Eq:Delta-ACE2-0}) as
follows:
\begin{equation}  \label{Eq:Delta-ACE2-ACE0}
\Delta
A_\text{CE}
=\epsilon_q^G\,A_{\text{CE},q}^G+\epsilon_g^G\,A_{\text{CE},g}^G
- \epsilon_q^{S}\, A_{\text{CE}}^{S}.
\end{equation}
In Eq.~(\ref{Eq:Delta-ACE2-ACE0}), the notations are: $A_{\text{CE},q}^G\equiv
2\,{z^*}^5+\frac{5}{2}\,z^*(1-{z^*}^2)-1$;
$A_{\text{CE},g}^G\equiv\frac{1}{2}\,{z^*}(5-{z^*}^4)-1$; and
$A_{\text{CE}}^{S}=2\,z^*-1$. We reconsider the numerical example of
Fig.~\ref{ACE_vs_z}, and note that around the chosen value $z^*=0.5$ the gluon
fusion subprocess largely dominates the deviation of
Eq.~(\ref{Eq:Delta-ACE2-ACE0}), due to $A_{\text{CE},g}^G\gg
A_{\text{CE},q}^G$,$A_{\text{CE}}^S$.  Actually, it is the only contribution
at $z^*=0.5$, because of the vanishing $A_{\text{CE},q}^G=A_{\text{CE}}^S=0$
at this point. This feature is found to hold more generally, also for the
other values of $M_G$ and $c$ different from those in Fig.~\ref{ACE_vs_z} or,
in other words, this choice is optimal in the sense that $A_{\rm CE}$ shows
maximal sensitivity to RS paramenters there.

\par To get an ``estimator'' that determines the spin-2 parameter space where
the spin-0 hypothesis could be excluded, the deviation
(\ref{Eq:Delta-ACE2-ACE0}) should be compared with the statistical uncertainty
on $A_\text{CE}$ expressed in terms of the desired number ($k$) of standard
deviations. We have the condition
\begin{equation}  \label{Eq:StandDev}
\vert\Delta
A_\text{CE}\vert=k\cdot\delta A_\text{CE},
\end{equation}
where, taking into account that numerically $(A_\text{CE}^G)^2\ll
1$ at $z^*\simeq 0.5$,
\begin{equation} \label{Eq:stat}
\delta A_\text{CE}
=\sqrt{\frac{1-{(A_\text{CE}^G)}^2}{N_{\rm{min}}}}
\approx \sqrt{\frac{1} {N_{\rm{min}}}}.
\end{equation}
From Eqs.~(\ref{Eq:StandDev}) and (\ref{Eq:stat}), one therefore obtains
\begin{equation} \label{Eq:SPIN0-EXCL}
N_{\rm{min}}=N_{\rm{min}}^S
\approx\left(\frac{k}{\epsilon_g^G\,A_{\text{CE},g}^G}\right)^2.
\end{equation}

\par Fig.~\ref{epsilongg} shows the contribution of gluon-gluon fusion to
graviton production at the LHC as a function of $M_G$.  One finds that, for
$c=0.1$, the SM background contribution is less than $1\%$ for all values of
$M_G$ considered.\footnote{Conversely, the figure shows that, for $c=0.01$,
this background can be appreciably higher, see also Table~\ref{tab:identify},
and should be taken into account.}  Extracting the value of $\epsilon_g^G$
from Fig.~\ref{epsilongg}, one can easily evaluate from
Eq.~(\ref{Eq:SPIN0-EXCL}) the minimal number of event samples required to
exclude the spin-0 hypothesis (hence to establish the spin-2). For example,
for $M_G=2$ TeV and $c=0.1$, we would find that $N_{\rm min}^{S}\simeq 38$ at
the $1\,\sigma$ level ($k=1$), compatible with results in
Ref.~\cite{Cousins:2005pq}. The limiting number $N_{\rm min}^{S}$ required for
identification against spin-0 smoothly increases with $M_G$ as $\epsilon_g^G$
decreases, as shown in Fig.~\ref{epsilongg}.

\par
\begin{figure}[bht] 
\centerline{ \hspace*{-1.0cm}
\includegraphics[width=12.0cm,angle=0]{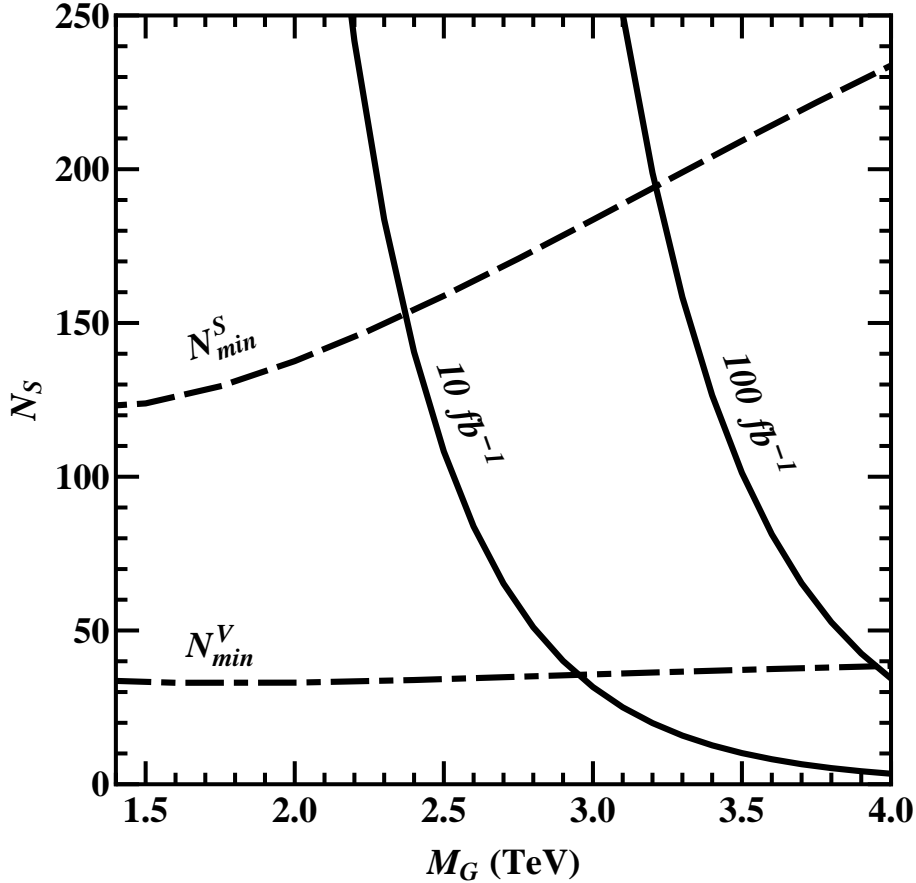}}
\vspace*{0.0cm} 
\caption{\label{Nmin} Minimal numbers of spin-2 events required to
discriminate the spin-2 from the spin-0 ($N_\text{min}^S$) and spin-1
($N_\text{min}^V$) hypotheses at 95\% C.L. as a function of $M_G$ at
$c=0.1$. The steep solid lines represent the number of spin-2 resonance
(signal) events, $N_S$, which could be measured at the LHC at two chosen
values of the luminosity, $\Lumint=10$ fb$^{-1}$ and $100$ fb$^{-1}$,
respectively.}
\end{figure}

\begin{figure}[bht] 
\centerline{ \hspace*{-0.0cm}
\includegraphics[width=12.0cm,angle=0]{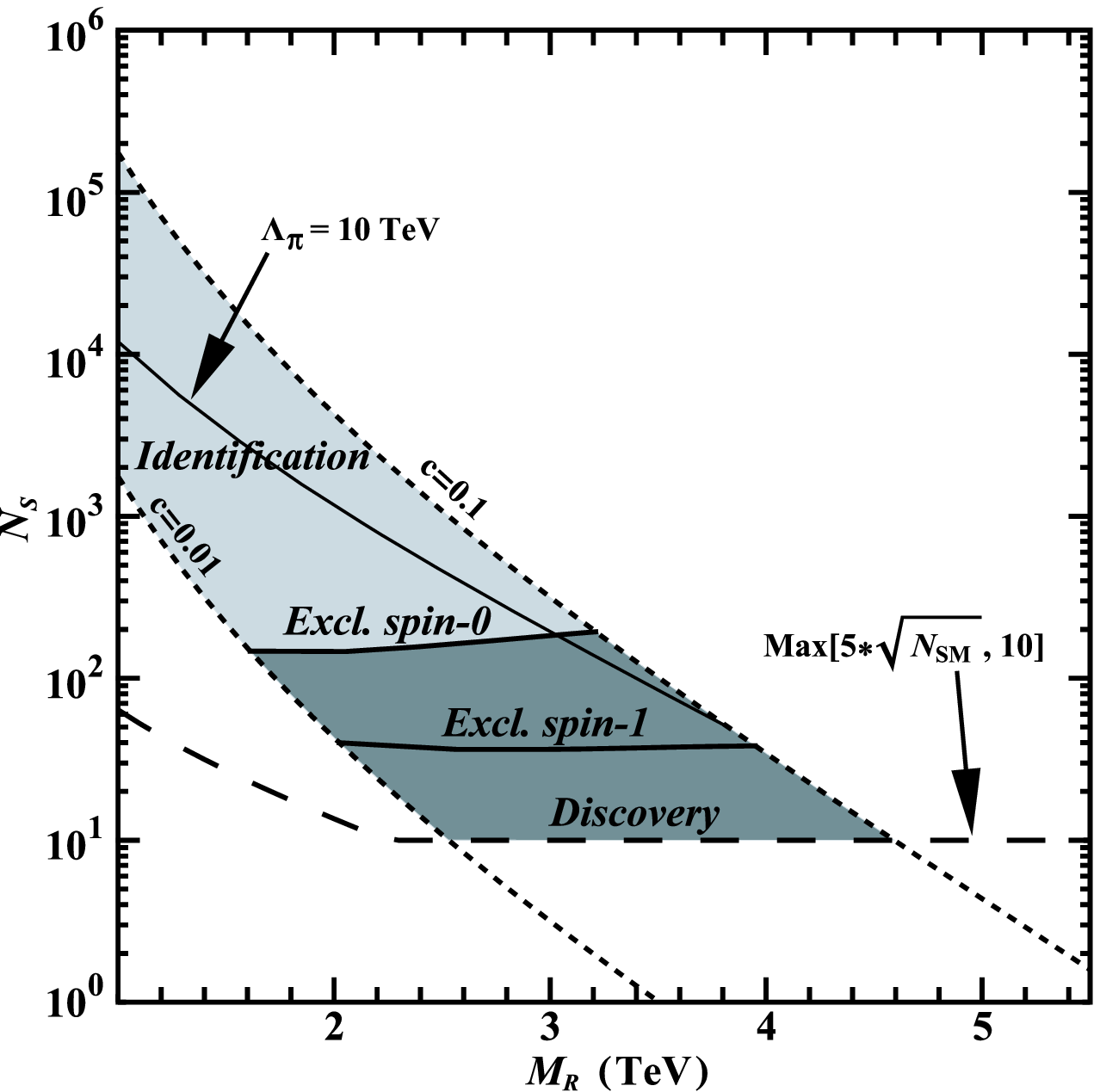}}
\vspace*{-0cm} 
\caption{\label{Ns_vs_MR_excl_id} Same as in Fig.~\ref{Ns_vs_MG} but with
exclusion limits and identification reach at 95\% C.L.\ and ${\cal L}_{\rm
int}=100$ fb$^{-1}$. The channels $l=e,\mu$ are combined. The theoretically
favored region, limited by the $\Lambda_\pi=10~\text{TeV}$ and $c=0.1$ lines,
is also indicated.}
\end{figure}

 The behavior of $N_{\rm min}^{S}$ {\it vs.}\ $M_G$ is presented for
$c=0.1$ in Fig.~\ref{Nmin}. It is derived from the full calculation including
detector cuts, using the general Eq.~(\ref{Eq:StandDev}), with
$k=\sqrt{3.84}=1.96$, corresponding to the exclusion of the spin-0 resonance
at 95\% C.L. The two solid lines in this figure represent the number of
resonance signal events $N_S$ {\it vs}. $M_G$ at luminosities $\Lumint=10$
fb$^{-1}$ and $100$ fb$^{-1}$, respectively. Their intersections with the line
of $N_{\rm min}^S$ {\it vs.}  $M_G$ determine the value of the graviton mass
where the spin-0 hypothesis can be excluded. In this example, $M_G=2.4$ and
$3.2$~TeV at $\Lumint=10$ and at $100$ fb$^{-1}$, respectively. By repeating
this procedure for all other allowed values of the parameter $c$, from 0.1
down to 0.01, one can determine the corresponding values of $N_{\rm min}$ for
excluding the spin-0 hypothesis and the related values of $M_G$. The result
for $N_{\rm min}$, at integrated LHC luminosity of 100 $\text{fb}^{-1}$, is
displayed in Fig.~\ref{Ns_vs_MR_excl_id} as a function of graviton mass $M_G$
(95\% C.L.). The resulting domain defines the identification reach on the
spin-2 of the resonance.

\par Quite similarly, one can separately estimate the minimal number of events
required to discriminate the spin-2 from the spin-1 hypotheses, on the basis
of the difference in Eq.~(\ref{Eq:Delta-ACE2-1}). In the approximations
adopted above for the spin-0 case, one would get the expression
\begin{equation} \label{Eq:SPIN1-EXCL}
N_{\rm{min}}\equiv
N_{\rm{min}}^V
\approx\left(\frac{k}{\epsilon_g^G\,A_{\text{CE},g}^G-A_{\rm CE}^V}\right)^2.
\end{equation}
From Eq.~(\ref{Eq:SPIN1-EXCL}) and Fig.~\ref{epsilongg}, the minimal number of
spin-2 RS events needed for excluding the spin-1 hypothesis, and the relevant
resonance masses $M_G$ depending from the LHC luminosity, can be obtained. The
example for $c=0.1$ is reported in Fig.~\ref{Nmin}.

\par The results of the full calculations of $A_{\rm CE}(M_R)$ including
detector cuts, to determine the exclusions of the spin-1 and the spin-0
hypotheses in terms of the corresponding minimal number of events $N_{\rm
min}^{V}$ and $N_{\rm min}^{S}$, respectively, are presented for ${\cal
L}_{\rm int}=100$ fb$^{-1}$ in Fig.~\ref{Ns_vs_MR_excl_id}. This figure is
conceptually similar to Fig.~\ref{Ns_vs_MG}. The combination of the spin-0 and
spin-1 rejection domains determines a common exclusion area and thus the
domain where the spin-2 of the RS graviton resonance can be established. This
is the light gray domain labelled ``Identification''.

\par Equivalent to the above procedure,
the identification $M_G$ and $c$ can be assessed by means of a
``conventional'' (and simple minded) $\chi^2$ criterion, where the $\chi^2$
function is defined as:
\begin{equation} \label{Eq:five-three}
\chi^2=\left[\frac{\Delta A_\text{CE}} {\delta
A_\text{CE}}\right]^2,
\end{equation}
with $\Delta A_{\rm CE}$ represented by Eqs.~(\ref{Eq:Delta-ACE2-0}) and
(\ref{Eq:Delta-ACE2-1}) to obtain the exclusion domains of the spin-0 and
spin-1 hypotheses, respectively, and the statistical uncertainty
\begin{equation} \label{Eq:StatUncert}
\delta A_\text{CE}=\sqrt{\frac{1-{(A_\text{CE}^G)}^2}
{\epsilon_l {\cal L}_\text{int}\sigma(G_{ll})}}.
\end{equation}
Like before, the RS model can be assumed to be the ``true'' one, and the (95\%
C.L.) exclusion domains of spin-0 and spin-1 can be determined by requiring
$\chi^2=3.84$, as pertinent to a one-parameter fit.\footnote{The parameters
$M_R$ and $c$ are constrained via Eq.~(\ref{Eq:kappa-def}), rendering a
two-parameter constraint effectively a one-parameter constraint.}  Like
before, the maximal sensitivity of $A_{\rm CE}$ to the spin-2 RS resonance
parameters is generally achieved for $z^*=0.5$. Again, we combine the channels
$l=e,\mu$. The 95\% C.L. identification reach of the spin-2 hypothesis in the
$(M_G, c)$ plane then results from the domain complementary to the
combination of the spin-0 and spin-1 exclusion domains.  In fact, the spin-0
exclusion is more restrictive than that for spin-1, as discussed above.
\section{Results in the RS parameter plane}
\label{sect:results}
\begin{figure}[htb] 
\centerline{ 
\includegraphics[width=14.0cm,angle=0]{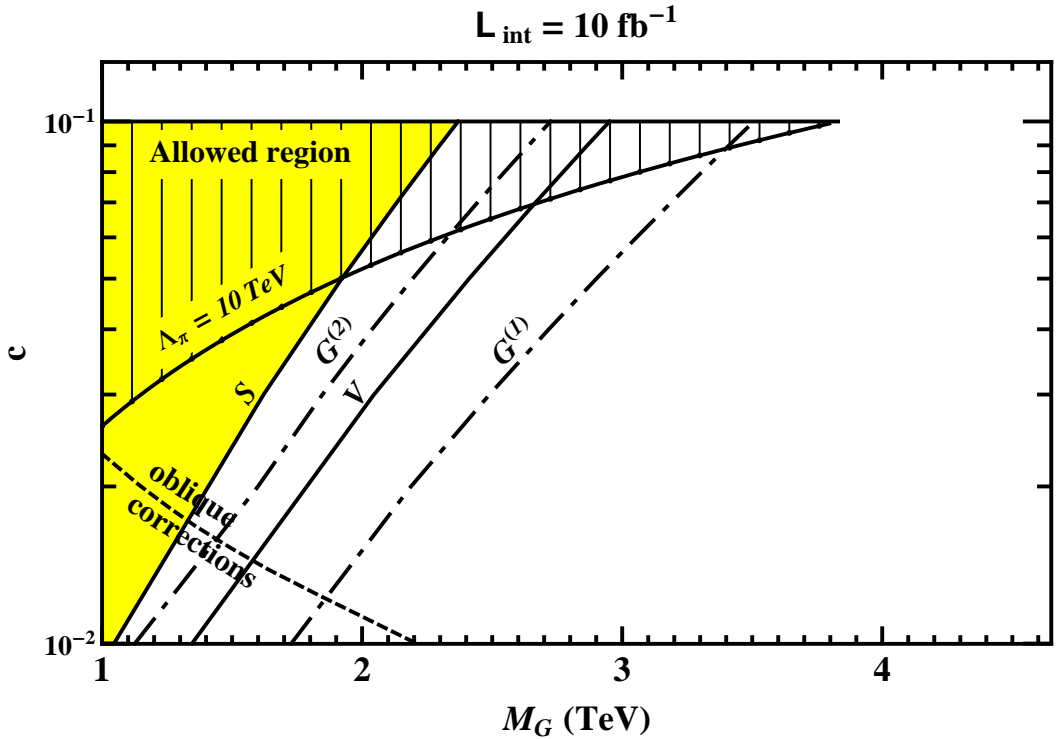}}
\caption{\label{id_vs_MG10fb} Discovery limits ($G^{(1)}$ and $G^{(2)}$,
$5\sigma$ level) and identification reaches ($V$, $S$, 95\% C.L.)  on the
spin-2 graviton parameters in the plane ($M_G$, $c$), using the lepton pair
production cross section and center--edge asymmetry, at the LHC with
integrated luminosity of 10 fb$^{-1}$.  The theoretically favored region,
$\Lambda_\pi<10~\text{TeV}$ (hatched), and bounds from the global fit to the
oblique parameters, are also indicated.}
\end{figure}

\begin{figure}[htb] 
\centerline{ 
\includegraphics[width=14.0cm,angle=0]{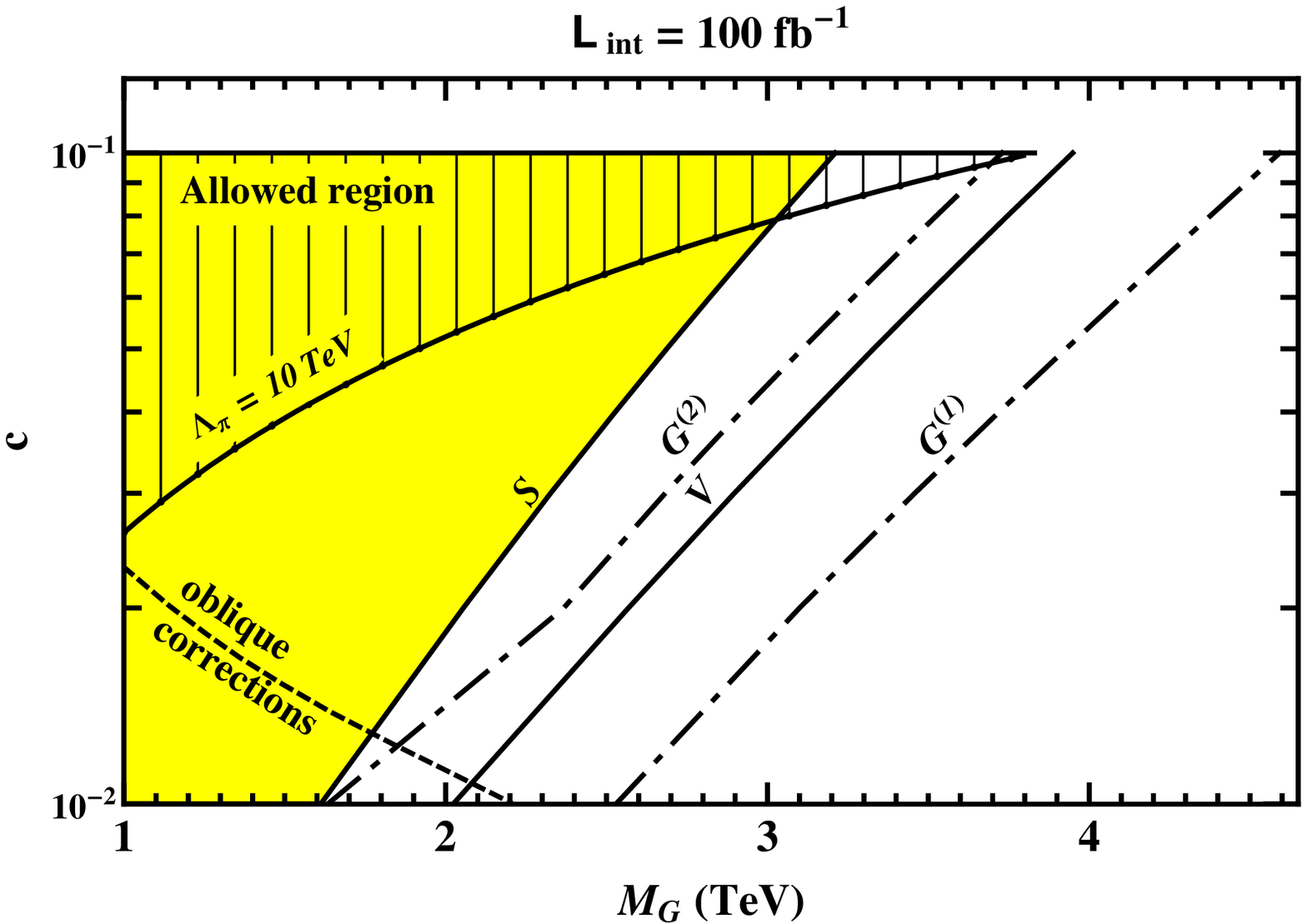}}
\caption{\label{id_vs_MG100fb} Same as in Fig.~\ref{id_vs_MG10fb} but
for the LHC integrated luminosity of 100 fb$^{-1}$.}
\end{figure}

The combined results of the previous section are presented, in the RS
parameter plane $(M_G,c)$, in Figs.~\ref{id_vs_MG10fb}
and~\ref{id_vs_MG100fb}, for LHC integrated luminosities of 10 fb$^{-1}$
and 100 fb$^{-1}$, respectively. The solid lines in
these figures with attached labels ``$S$'' and ``$V$'' represent exclusion
limits at the 95\% C.L.

\par The dot-long-dashed line labelled as ``$G^{(1)}$'' represents the
$5\sigma$ discovery reach on the lowest-lying RS graviton $G\equiv
G^{(1)}$ at each assumed luminosity. The resonance can be discovered if its
representative point ($M_G,c$) lies to the left of this curve. Conversely, $G$
could not be discovered if the corresponding representative point lies to the
right.\footnote{Of course, all such statements must be understood in a
statistical sense, as specified by the confidence level.}

\par According to the discussion in Sec.~\ref{sect:RSidentification}, the
domain to the left of the line labelled ``$V$'' represents the ($M_G,c$)
values for which the spin-1 hypothesis for the first RS resonance can be
excluded, but the spin-0 hypothesis is still left open. Finally, the domain to
the left of the line labelled as ``$S$'', represents the ($M_G,c$) values of
the first RS resonance for which the alternative spin-0 hypothesis can be
excluded. Reflecting the results in Sec.~\ref{sect:RSidentification}, this
solid line lies to the left of the ``$V$'' one. Therefore, we may assume the
combination of the two respective domains shown as the shaded area, where both
spin-1 and spin-0 hypotheses are excluded, to represent the lowest-lying RS
resonance ``identification'' domain where the spin-2 character can be fully
discriminated.\footnote{Of course, barring other possibilities for the spin of
the discovered resonance, not considered as basic starting point in our
discussion.}

\par Numerically, one can read from Fig.~\ref{id_vs_MG10fb} that, for
$\Lumint=10$~fb$^{-1}$ and the coupling $c$ in the theoretically favored
range, spin-1 can be excluded up to $M_G=1.3$~TeV for $c=0.01$, and up to
$M_G=2.9$~TeV at $c=0.1$. Moreover, the spin-2 character of the resonance can
be identified by spin-0 exclusion up to $M_G=1.0$~TeV for $c=0.01$ and up to
$M_G=2.4$~TeV for $c=0.1$. At the higher luminosity, $\Lumint=100$~fb$^{-1}$,
Fig.~\ref{id_vs_MG100fb} indicates that the spin-1 hypothesis may be
excluded up to $M_G=2.0$~TeV for $c=0.01$ and $M_G=4.0$~TeV for $c=0.1$. 
The spin-2 of the RS resonance can be identified by spin-0 exclusion up to
$M_G=1.6$~TeV for $c=0.01$ and $M_G=3.2$~TeV for $c=0.1$.

Figs.~\ref{id_vs_MG10fb} and~\ref{id_vs_MG100fb} show rather clearly how the
request of discriminating the spin-2 resonance from {\it both} the spin-1 and
spin-0 hypotheses substantially reduces the ``allowed'' discovery domain in
the ($M_G,c$) plane.  Furthermore, this request reduces the size of the domain
that would be allowed by the weaker condition of only discriminating spin-2
from the spin-1 hypothesis by a non-negligible amount.  The theoretically
``preferred'' region, bounded from below (in $c$) by the line
$\Lambda_\pi\lsim 10$ TeV and also represented in these figures, is the source
of a further, dramatical, restriction of the allowed region. However, this
bound, rather than literally, should be considered as an order of magnitude
indication. Also, the bound from the global fit to the oblique parameters,
taken from Refs.~\cite{Davoudiasl:2000jd,han2000}, has a qualitative
character.


If producing more than one resonance is kinematically (and statistically)
feasible, the fact that in the RS model the excitation spacing is proportional
to the root of the $J_1$ Bessel function also provides a signal for this
scenario. It should therefore be important to examine the probability of
observing the second excitation if the first resonance is
discovered. Quantitatively, one should evaluate the cross section times the
leptonic branching fraction for the DY production {\it via} the second
excitation of the graviton state, $G^{(2)}$, as a function of the lowest-lying
graviton mass $M_G$, and for the parameter $c$ ranging from 0.1 down to
0.01. Neglecting interference between the two resonances, the result is that
the $n=2$ graviton state $G^{(2)}$ can be discovered at the LHC with 10
fb$^{-1}$ of integrated luminosity, if the mass of the {\it lowest}-lying one
$G\equiv G^{(1)}$ is less than 1.1~TeV (2.7 TeV) at $c=0.01$
($c=0.1$). With 100 fb$^{-1}$, discovery of $G^{(2)}$ is possible if
the mass $M_G$ of the first excitation $G^{(1)}$ is less than 1.6 TeV (3.7
TeV) at $c=0.01$ ($c=0.1$).  
All these numbers are at the $5\sigma$ level.
This clearly represents a significant discovery reach.
The criterion to assess the discovery reach on $G^{(2)}$ has been the same 
as for the discovery of $G^{(1)}$.

\par In Figs.~\ref{id_vs_MG10fb} and~\ref{id_vs_MG100fb}, the line labelled as
``$G^{(2)}$'', represents the values of the first resonance mass $M_G$ and $c$
for which the second state, $G^{(2)}$, can also be discovered. In the
($M_G,c$) domain for $G$ located to the right of that line, $G^{(2)}$ could
not be discovered. Conversely, for graviton $G$ in the domain to the left of
that line, the second graviton $G^{(2)}$ can also be discovered. One can
notice that in both figures this line is located between the ``$V$'' and the
``$S$'' lines, and this seems to be a general feature.

\par Therefore, taking into account the above discussion and the meaning of
the various lines in Figs.~\ref{id_vs_MG10fb}
and~\ref{id_vs_MG100fb}, one can envisage the following possible
scenarios in the discovery and identification of the lowest-lying RS resonance:
\begin{itemize}
\item[(i)] $G$ is discovered in the strip between the ``$V$'' and
``$G^{(1)}$'' lines, in which case only discovery of $G$ is possible, but no
identification;
\item[(ii)] in the strip between the ``$G^{(2)}$'' and the ``$V$'' lines,
where angular analysis can be used to exclude spin-1, but no spin-0 rejection
and no production of the second resonance;
\item[(iii)] $G$ is found in the strip between the ``$S$'' and the
``$G^{(2)}$'' lines, in this case analysis based on the angular asymmetry
$A_\text{CE}$ can be performed to exclude spin-1, not spin-0 yet, but the
second resonance $G^{(2)}$ can be discovered and the RS spectrum test can be
performed;\footnote{In particular, as shown by Fig.~\ref{id_vs_MG100fb},
if one takes literally into account the severe bound $\Lambda_\pi\leq 10$~TeV,
at the high luminosity of 100~fb$^{-1}$ such mass spectrum test should be
operative in the full discovery region.}
\item[(iv)] in the region to the left of the ``$S$'' line indicated as the
shaded area, the spin-2 character of the RS lowest-lying resonance can be
identified and, in addition, the RS graviton mass spectrum can be verified by
the discovery of the second resonance. Thus, the model would be doubly tested,
by both the mass spectrum and the spin-2 angular analysis, and the RS
resonance $G$ clinched completely.
\end{itemize}
\section{Concluding remarks} 
\label{sect:conclusions}
In conclusion, we have considered the RS scenario, assuming a discovery can be
made in the form of a resonance in the dilepton cross section.  We have
determined by an $A_{\rm CE}$-based analysis up to what mass the spin-2
property can be established. This basically amounts to excluding the spin-0
hypothesis, in which case the spin-1 alternative will be automatically
excluded. Additionally, we point out that in the parameter space where this
spin determination is possible, the second resonance, with its characteristic
mass, is also visible.

\par The analysis for the identification of the graviton spin can be performed
using as basic observable the angular distribution of leptons itself, rather
than the normalized angular-integrated center-edge asymmetry $A_{\rm CE}$. Of
course, one might expect that, with extremely high statistics, the two
approaches could lead to equivalent results because, after all, the same data
are used. However, if maintaining the same level of (high) statistics in each
angular bin is required for the measurement of the differential cross section,
an advantage of the method exposed here should be that a comparatively smaller
event sample would be needed for the centre-edge asymmetry $A_{\rm CE}$ to
obtain the contraints on the RS resonance at a given C.L. In addition, besides
being ``transparent'' to spin-1 exchanges, the asymmetry $A_{\rm CE}$ might be
less sensitive to systematic uncertainties (hence more sensitive to RS
parameters) than an ``absolute'' angular distribution, because such
uncertainties can be hoped to (at least partially) cancel.

\par The systematic uncertainties, not yet accounted for in the analysis
presented above, originate from many sources, e.g., the accuracy of the
theoretical calculations, the differences in the phenomenological
determinations of the parton distribution functions, 
including the uncertainties in the cross section predictions related to 
the choice of factorization scales as extensively discussed in 
Ref.~\cite{Mathews:2004xp}, and the experimental
uncertainties on the acceptance, electron identification efficiency,
luminosity, and so on. 
Of course, the expectation is that these uncertainties should be mitigated 
by the basic observable $A_{\rm CE}$ being a ratio of (integrated) 
cross sections.
The dominant experimental systematic error on the event
rate can be expected to originate from the luminosity measurement, we
conservatively assume at the 10\% level, while uncertainties on efficiencies
and acceptances should be at (or below) the 1\% level
\cite{Allanach:2002gn}. Preliminary estimates seem to confirm the cancellation
of such systematic uncertainties in $A_{\rm CE}$ alluded to before, and their
almost negligible impact on the results for the identification reach on the
spin-2 RS resonances presented in Sec.~\ref{sect:results}. We plan to
investigate the effect of the other uncertainties mentioned above in a future
analysis.

\vspace{0.5cm} 
\leftline{\bf Acknowledgements} 
\par\noindent This research has been partially supported by the Abdus Salam
ICTP and the Belarusian Republican Foundation for Fundamental Research. 
AAP also acknowledges the support of MiUR (Italian Ministry of University and 
Research) and of Trieste University.
The work of PO has been supported by The Research Council of
Norway, and that of NP by funds of MiUR.


\end{document}